\documentclass[%
 reprint,
superscriptaddress,
 amsmath,amssymb,
 aps,
]{revtex4-2}

\usepackage{placeins}

\usepackage{graphicx}
\usepackage{dcolumn}
\usepackage{bm}
\usepackage{braket}
\usepackage{color}
\usepackage{amsthm}
\usepackage{dsfont}
\usepackage[absolute]{textpos}
\usepackage{xcolor}
\usepackage{fancyhdr}
\usepackage{listings}
\usepackage{algorithmic}
\usepackage[ruled,vlined]{algorithm2e}
\usepackage{listings}
\usepackage{xcolor}

\definecolor{codegreen}{rgb}{0,0.6,0}
\definecolor{codegray}{rgb}{0.5,0.5,0.5}
\definecolor{codepurple}{rgb}{0.58,0,0.82}
\definecolor{backcolour}{rgb}{0.95,0.95,0.92}

\lstdefinestyle{mystyle}{
    backgroundcolor=\color{backcolour},   
    commentstyle=\color{codegreen},
    keywordstyle=\color{magenta},
    numberstyle=\tiny\color{codegray},
    stringstyle=\color{codepurple},
    basicstyle=\ttfamily\footnotesize,
    breakatwhitespace=false,         
    breaklines=true,                 
    captionpos=b,                    
    keepspaces=true,                 
    numbers=left,                    
    numbersep=5pt,                  
    showspaces=false,                
    showstringspaces=false,
    showtabs=false,                  
    tabsize=2
}

\usepackage{appendix}

\preprint{APS/123-QED}

\begin{document}
\title{Dimension-adaptive machine-learning-based quantum state reconstruction}


\author{Sanjaya Lohani}
\email[]{slohan3@uic.edu}
\affiliation{Dept.~of Electrical \& Computer Engineering, University of Illinois Chicago, Chicago, IL 60607, USA}
\author{Sangita Regmi}
\affiliation{Dept.~of Electrical \& Computer Engineering, University of Illinois Chicago, Chicago, IL 60607, USA}
\author{Joseph M. Lukens}
\affiliation{Quantum Information Science Section, Oak Ridge National Laboratory, Oak Ridge, TN 37831, USA}
\author{\\Ryan~T. Glasser}
\affiliation{Tulane University, New Orleans, LA 70118, USA}
\author{Thomas~A. Searles}
\email[]{tsearles@uic.edu}
\affiliation{Dept.~of Electrical \& Computer Engineering, University of Illinois Chicago, Chicago, IL 60607, USA}
\author{Brian~T. Kirby}
\email[]{brian.t.kirby4.civ@army.mil}
\affiliation{Tulane University, New Orleans, LA 70118, USA}
\affiliation{DEVCOM Army Research Laboratory, Adelphi, MD 20783, USA}
\date{\today}

\begin{abstract}
We introduce an approach for performing quantum state reconstruction on systems of $n$ qubits using a machine-learning-based reconstruction system trained exclusively on $m$ qubits, where $m\geq n$. This approach removes the necessity of exactly matching the dimensionality of a system under consideration with the dimension of a model used for training. 
We demonstrate our technique by performing quantum state reconstruction on randomly sampled systems of one, two, and three qubits using machine-learning-based methods trained exclusively on systems containing at least one additional qubit.
The reconstruction time required for machine-learning-based methods scales significantly more favorably than the training time; 
hence this technique can offer an overall savings of resources by leveraging a single neural network for dimension-variable state reconstruction, obviating the need to train dedicated machine-learning systems for each Hilbert space.
\end{abstract}

\flushbottom
\maketitle
\begin{textblock}{13.3}(1.4,15)
\noindent\fontsize{7}{7}\selectfont \textcolor{black!30}{This manuscript has been co-authored by UT-Battelle, LLC, under contract DE-AC05-00OR22725 with the US Department of Energy (DOE). The US government retains and the publisher, by accepting the article for publication, acknowledges that the US government retains a nonexclusive, paid-up, irrevocable, worldwide license to publish or reproduce the published form of this manuscript, or allow others to do so, for US government purposes. DOE will provide public access to these results of federally sponsored research in accordance with the DOE Public Access Plan (http://energy.gov/downloads/doe-public-access-plan).}
\end{textblock}
%
%
\thispagestyle{empty}

\section{Introduction}

Estimating the properties of a quantum system through measurement is a task of fundamental importance in quantum information science. Although methods exist for the partial characterization of quantum systems requiring relatively few measurements~\cite{ekert2002direct, Flammia2011, Spengler2012, huang2020predicting, Eisert2020, lukens2021bayesian}, the complete reconstruction of a density matrix has the distinct advantage of providing full information on any property of the system. Complete reconstruction requires quantum state tomography (QST), where repeated measurements on an ensemble of identically prepared systems are used to estimate the system's density matrix. In general, QST consists of the preparation and experimental collection of measurement results and the purely classical and computational step of recovering the density matrix most consistent with the measurement results, known as quantum state reconstruction \cite{thew2002qudit,altepeter2005photonic,james2005measurement}. Various methods exist for performing quantum state reconstruction, including maximum likelihood estimation~\cite{Hradil1997, Banaszek1999, James2001, Lvovsky2004, altepeter2005photonic, james2005measurement, smolin_efficient_2012}, Bayesian inference \cite{Blume2010, Huszar2012, Kravtsov2013, Seah2015,Granade2016, Williams2017,Mai2017, lukens2020practical, simmerman2020efficient, lukens2021bayesian, Lu2021, chapman2022}, and machine-learning-based techniques -- supervised learning \cite{lu2018separability,lohani2020machine,danaci2021machine,ahmed2021classification,lohani2021experimental,lohani2021improving,lohani2022data,torlai2018neural,torlai2019integrating,melkani2020eigenstate,hsieh2022direct,genois2021quantum,teo2021benchmarking,tiunov2020experimental,palmieri2020experimental,neugebauer2020neural, wang2021quantum}, semi-supervised learning \cite{ahmed2021quantum, carrasquilla2019reconstructing, lohani2020generative}, and reinforcement learning~\cite{borah2021measurement}.



The exponential scaling of Hilbert space dimension with the number of qubits presents a challenge both experimentally and computationally for implementations of QST. 
For QST to reconstruct an arbitrary density matrix with low uncertainty, the measured bases should ideally span the entire Hilbert space. 
Hence, the number of distinct measurement bases desired will always scale exponentially. Similarly, the computational cost required to perform quantum state reconstruction using most leading techniques, such as maximum likelihood or Bayesian estimation, also scales exponentially with system size. While the resources required to perform the experimental measurements required for QST often eclipse the reconstruction time for small quantum systems (e.g., one or two qubits), this is not necessarily the case for larger quantum systems \cite{gross2010quantum,haffner2005scalable,Lu2021}. For this reason, significant research has focused on developing alternative quantum state reconstruction methods with more favorable computational scaling.

One recently proposed approach for confronting quantum state reconstruction's  computational cost is to frontload the exponential scaling into the training period of a machine-learning-based system \cite{lohani2020machine,lohani2021experimental,danaci2021machine}. In particular, recent results applying pre-trained networks to near-term intermediate scale quantum (NISQ) devices of up to four qubits revealed a significant increase in the training time as a function of the dimension of the underlying space, but only an extremely modest increase in the reconstruction time~\cite{lohani2021experimental}. For example, after training, the reconstruction time was 0.77~ms for single-qubit systems, rising only to 0.8~ms for four qubits---a near-trivial increase considering the eight-fold growth in Hilbert space dimension. 
Note that the training of a network only needs to be performed once, and subsequently the network can be applied to any future datasets using comparatively modest resources. For example, as described above, the pre-trained network for four-qubit systems can always perform full state reconstruction in 0.8~ms (on the hardware used in~\cite{lohani2021experimental}) without any additional training.

\begin{figure*}[ht]
\centering
\includegraphics[width=.85\linewidth]{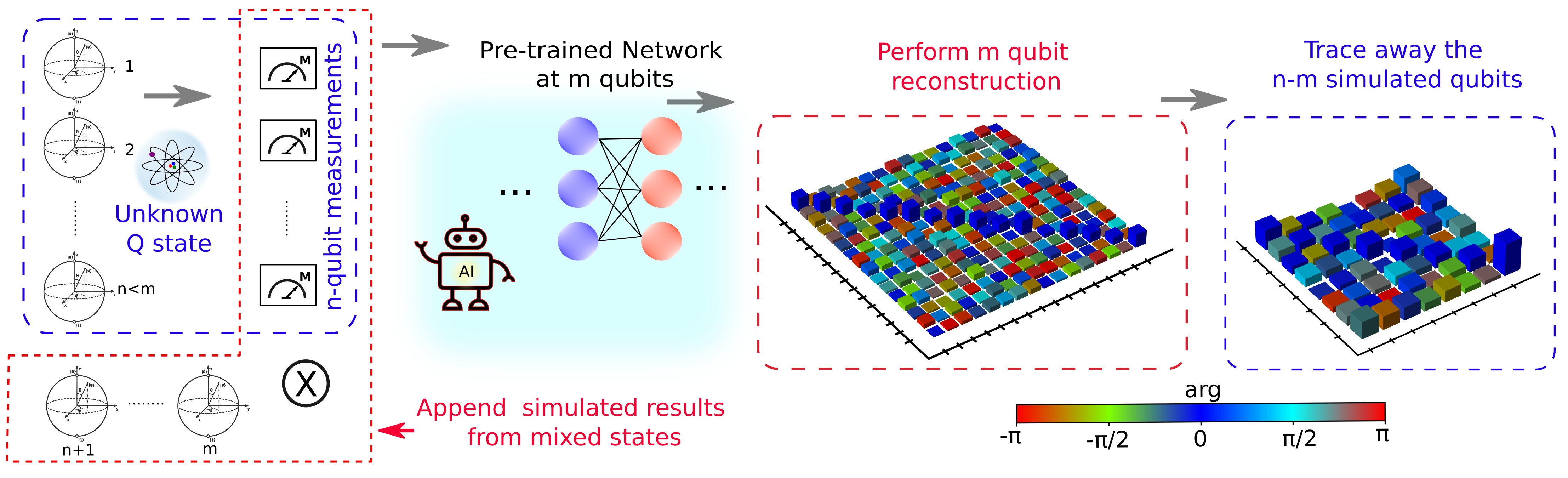}
\caption{A schematic of our approach for dimension-adaptive quantum state reconstruction. The estimation of $n$-qubit quantum states uses a machine-learning-based reconstruction system trained on $m$ qubits, where $m\geq n$. First, we append virtual results for $m\,-\,n$ qubits via engineered padding (red dotted box) to $n$-qubit measurements (blue dotted box), and feed the padded measurements into a network pre-trained for $m$ qubits. At the output, the network returns an $m$-qubit density matrix, which is partially traced to return an estimate of the unknown quantum system of $n$ qubits.}
\label{fig:intro}
\end{figure*}

Although machine-learning-based reconstruction systems can be trained over the entire Hilbert space and used to reconstruct arbitrarily mixed states, the training generally focuses on a fixed Hilbert space dimension in order to limit the number of trainable parameters required to describe the system. In other words, while it is in principle possible to train a network to accept variable-dimension input states, it requires a dramatic, and often practically infeasible, increase in network size and training time. 
Therefore, an existing weakness of this approach is that a given trained network can only be applied to systems of precisely the dimension on which it was trained and does not generalize to smaller Hilbert spaces, instead requiring a separate system to be trained for every dimension.

Here we address this current limitation by proposing an approach for quantum state reconstruction on systems of $n$ qubits using a machine-learning-based system trained on $m$ qubits, where $m\geq n$. We begin by generally relating the average reconstruction fidelity of an $m$-qubit quantum state to the average reconstruction fidelity of any of its reduced density matrices by applying the monotonicity of the fidelity. We then extract this relationship specifically for ensembles of states randomly sampled according to the Hilbert--Schmidt (HS) measure for $m\in\{2,3,4\}$. We interpret these results to indicate that reconstruction systems intended explicitly for $m$-qubit reconstruction implicitly inherit the capacity to perform $n<m$ qubit reconstructions.  In addition, we describe a method, pictured in Fig.~\ref{fig:intro}, for augmenting $n<m$ qubit tomography measurement results to $m$ qubits such that the desired reconstruction can be obtained through the partial trace. Finally, we demonstrate our approach using simulated tomographic measurement data for $n\leq m$ qubits, 
complete inference utilizing networks trained with $m\in\{2,3,4\}$, and discuss the performance of our method.

\section{Machine-Learning-Based Quantum State Reconstruction}
\label{sec:ML}

In this section, we describe the general details of our machine-learning-based approach to $m$-qubit QST. We implement a convolutional neural network (CNN) with a convolutional unit of kernel size (2, 2), strides of 1, ReLU as an activation function, and 25 filters. The output of the CNN is fed into the next layer, which performs pooling with a pool-size (2, 2), followed by a second convolutional unit of the same configuration. Then, we combine two dense layers, followed by a dropout layer with a rate of 0.5, which is then attached to an output layer predicting $\tau$-vectors (Cholesky coefficients of the density matrix \cite{altepeter2005photonic}). The mean square loss between the target and predicted $\tau$ is evaluated and fed back to optimize the network's trainable parameters using the Adagrad optimizer. We use a learning rate of 0.01 and batch size of 100 for up to 300 epochs to train the network.
At the output layer is attached a pipeline that rearranges the predicted $\tau$-vectors into density matrices. The pipeline is built into the same graph of the network for the purposes of computing the average fidelity per epoch for cross-validation and outputting the density matrix directly to avoid post-processing. As an example, in the two-qubit case, the predicted $\tau$-vectors are rearranged to lower triangular matrices, $T$, expressed as
\begin{equation}
\begin{aligned}
    &T\,=\,
        \begin{bmatrix}
        \tau_0 & 0& 0& 0\\
        \tau_4+i\tau_5 & \tau_1 &0 &0\\
        \tau_{10}+i\tau_{11} & \tau_6+i\tau_7 &\tau_2 &0\\
        \tau_{14}+i\tau_{15} & \tau_{12}+i\tau_{13} &\tau_8+i\tau_9 &\tau_3\\
        \end{bmatrix}. \\
\end{aligned}
\label{eqn:lower_t}
\end{equation}
The density matrices follow as $\tilde{\rho}=\frac{TT^\dagger}{\textrm{Tr}(TT^\dagger)}$. Note that the physicality of $\tilde{\rho}$ is guaranteed through the Cholesky decomposition, which ensures positive semidefiniteness~\cite{James2001, altepeter2005photonic}. Finally, at the end of the network, the fidelity $F$ between the predicted density matrix $\tilde{\rho}$ and the target $\rho$ is evaluated as $F = \Big|\textrm{Tr}\sqrt{\sqrt{\tilde{\rho}}\rho\sqrt{\tilde{\rho}}}\Big|^2$. An in-depth description of the network architecture is given in \cite{lohani2021experimental}.

Previous work suggests an approximately exponential separation in the computational resources required to train a network of the type described above compared to using it for reconstruction. For example, an analysis of the explicit training and inference times for systems of one to four qubits showed that, using the same computational resources, the training of a one-qubit network took 123~s but only 0.77~ms to perform reconstruction, compared to 1380~s and 0.80~ms, respectively, for a four-qubit system \cite{lohani2021experimental}. The modest scaling in inference times is the main appeal of machine-learning-based reconstruction methods but comes at the cost of an intensive upfront training period. Such unfavorable scaling in the training times of neural networks are reminiscent of those found in reconstruction approaches based on maximum likelihood~\cite{gross2010quantum, haffner2005scalable} or Bayesian estimation~\cite{lukens2020practical, Lu2021}; however, the network has the advantage that these resources can be expended ahead of time and only once. 

Here we seek to further mitigate the overhead required for training by repurposing a network trained on systems of a particular dimension for inference of all lower-dimensional systems as well. The general approach is pictured in Fig.~\ref{fig:intro} and described in detail in Sec.~\ref{sec:extending}. In order to illustrate the proof of concept and train and test our reconstruction approach, we use mixed quantum states sampled according to the HS measure. The choice of sampling according to the HS measure as opposed to others is due to the unique property that it induces a flat Euclidean geometry into the mixed states \cite{sommers2003bures,zyczkowski2003hilbert} and has hence found wide adoption in various studies of quantum states. We note, however, that many other distributions of random quantum states exist and have been studied in various contexts, including as prior distributions for Bayesian inference~\cite{Granade2016, Mai2017, lukens2020practical} and training sets for machine-learning-based reconstruction~\cite{lohani2021improving}. (For completeness, in Appendix~\ref{APP:Bures} we reproduce the results of this manuscript using density matrices sampled according to the Bures metric, another distribution of longstanding significance in quantum information~\cite{sommers2003bures, al2010random}.)

To train the network, we sample 35,500 random quantum states $\rho$ according to the HS measure for the given $m$. 
We simulate the associated $6^m$ Pauli measurement outcomes for systems with $m\in\{2,3,4\}$ qubits directly from expectation values, which physically corresponds to the infinite-measurement limit (i.e., no statistical noise).
For each scenario, we split the sampled data into a training set comprising 35,000 states and a validation set of 500 states to cross-validate the network performance per epoch. 
After training, we generate test sets
that are entirely unknown to the trained network. The code to generate all datasets can be found in~\cite{LOHANI_machine-learning-for-physical-sciences_2022}.



\section{Reduced density matrix fidelity}
\label{sec:reduced}

Machine-learning-based techniques for quantum state reconstruction have been applied to systems of a variety of dimensions.  
Despite attaining high average reconstruction fidelity for the overall state, to our knowledge the way in which this translates to the fidelity of the reduced density matrices has not been considered.
Such an analysis is useful if, after tomography and reconstruction, study of a specific subspace is desired without performing additional reconstruction.

A relationship between the fidelity of two density matrices and the fidelity between any of their corresponding reduced density matrices follows immediately from the well-known property of monotonicity~\cite{nielsen1996entanglement,wilde2011classical}.
In particular, the fidelity $F(\rho_{AB},\sigma_{AB})$ between any two density matrices $\rho_{AB}$ and $\sigma_{AB}$, and corresponding reduced density matrices $\rho_{A}=\text{Tr}_{B}(\rho_{AB})$ and $\sigma_{A}=\text{Tr}_{B}(\sigma_{AB})$ is bounded by
$F(\rho_{AB},\sigma_{AB})\le F(\rho_{A},\sigma_{A})$ where $A$ and $B$ denote arbitrary bipartitions of each state.
In the context of quantum state reconstruction, we can consider $\rho_{AB}$ as the actual ground truth state and $\sigma_{AB}$ as the reconstruction.
Hence, the average reconstruction fidelity for any corresponding reduced density matrices over $n$ qubits of an $m$-qubit reconstruction is lower bounded by the average $m$-qubit reconstruction fidelity.
Note that we are only able to apply the monotonicity of the partial trace to reconstruction methods that guarantee the physicality of the final density matrix, such as described in Sec.~\ref{sec:ML} through the Cholesky decomposition.

It is worth emphasizing that the monotonicity of the fidelity only applies to specific pairs of density matrices.
Yet in the context of machine-learning-based tomography, we are interested primarily in averages over distributions of quantum states, as we seek to establish bounds on tomographic performance that would apply to a variety of initially unknown density matrices. And as discovered previously, the average performance of machine-learning-based reconstruction techniques can be heavily dependent on the distribution of density matrices used to calculate the average \cite{lohani2021improving}.
Therefore, we stress that the mean reconstruction fidelity obtained during training only bounds the average $n<m$ reduced density matrix reconstruction fidelities (through monotonicity) when the test states are drawn from the same distribution, or more precisely, when the distribution of $n$-qubit test states is equal to the distribution resulting from tracing out $m-n$ qubits from the $m$-qubit training distribution.
In other words, we cannot use monotonicity to obtain a completely general lower bound only based on network performance during training, as the averages depend on the distribution from which the states are drawn during deployment.
Note, however, that this complication only occurs when trying to develop a lower bound \emph{before} deployment. 
Alternatively, suppose we have prior knowledge regarding the distribution of states for a given deployment scenario. In that case, we can sample this distribution using any pre-trained network and extract the lower bound specifically for this use case.
The development of custom distributions of random quantum states that mimic various general features of quantum systems could potentially limit the impact of mismatched training and test distributions in practice~\cite{lohani2022data}.

\begin{figure}[tb!]
    \centering
    \includegraphics[width=\linewidth]{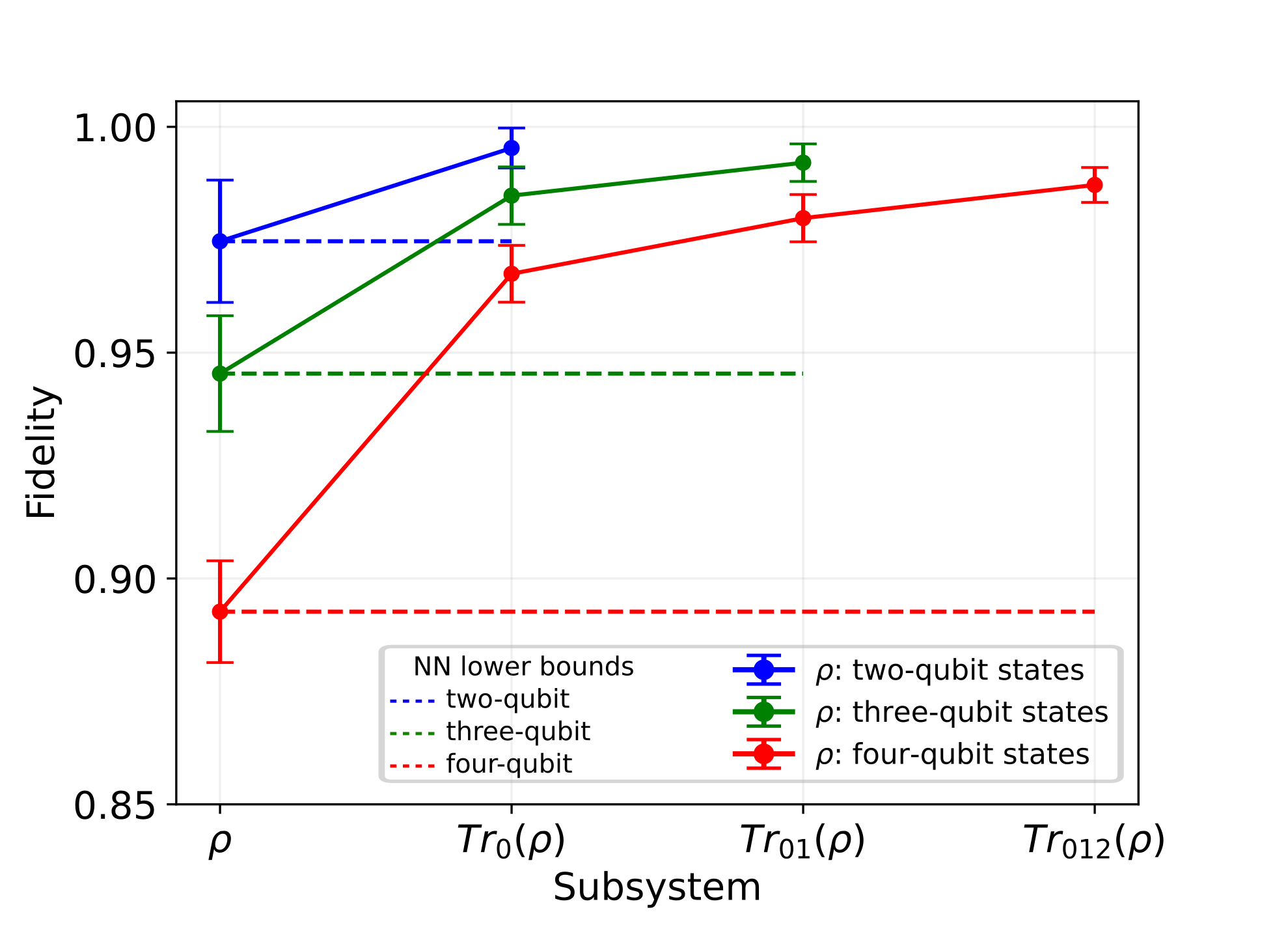}
    \caption{Reconstruction fidelity versus subsystem of predicted density matrix. For example, when $\rho$ represents a four-qubit system, then the subsystems $\text{Tr}_0(\rho)$, $\text{Tr}_{01}(\rho)$, $\text{Tr}_{012}(\rho)$ represent a three-qubit, two-qubit, and one-qubit quantum system, respectively. The dotted lines indicate the average reconstruction fidelity for an $m$-qubit system reconstructed with a network pre-trained on $m$ qubits.}
    \label{fig:fids_sub_system_all_HS}
\end{figure}

For illustrative purposes we now perform numerical simulations to compare the actual average reconstruction fidelity of the reduced density matrices to the lower bound determined by the monotonicity, using the methods described in Sec.~\ref{sec:ML} for $m\in\{2,3,4\}$ qubits.
The average reconstruction fidelity of each network was determined, which as described above, should serve as the lower bound on the average reconstruction fidelity of the reduced density matrices, and is plotted in Fig.~\ref{fig:fids_sub_system_all_HS} as the horizontal dashed lines.
We then use each network to reconstruct another, independently and generally different, ensemble of random quantum states sampled according to the HS metric on $m$ qubits, and for each reconstruction also perform a local trace for every decrement of one qubit and calculate the fidelity against the ground truth state.
As evident in Fig.~\ref{fig:fids_sub_system_all_HS}, in all cases the average fidelity outperforms the lower bound found from the monotonicity. 
Intuitively, the high fidelity of the reduced density matrices suggests that any $m$-dimensional reconstruction method implicitly includes some ability to reconstruct $n<m$ dimensional systems as well, provided it can be harnessed in a consistent fashion. This observation forms the inspiration for the general dimension-adaptive reconstruction scheme described in detail below.

\section{Extending system dimension with synthetic measurement results}
\label{sec:extending}
In the previous section, inference was performed on true $m$-qubit states using $m$-qubit-trained neural networks. By tracing down  these larger $m$-qubit states \emph{post-inference}, states with $n<m$ qubits were obtained leading to fidelities between the inferred and ground truth subsystems that increased steadily---a finding in agreement with expectations from monotonicity. Now we look to build upon these ideas and address the more challenging and unexplored situation where the target quantum system consists of $n$ qubits and one has access to a reconstruction apparatus designed only for $m>n$ qubits, thus requiring some method to bridge the mismatched Hilbert spaces while maintaining high accuracy.
We approach this problem by constructing extensions of states of $n$ qubits to $m>n$ qubits through the use of simulated measurement results.
Even though, as discussed below, monotonicity does not necessarily apply to these situations, intuitively we are exploiting the results of Sec.~\ref{sec:reduced}: the reduced density matrices of a state reconstructed using machine-learning-based methods maintain or, as in Fig.~\ref{fig:fids_sub_system_all_HS}, improve fidelity in comparison with the reconstruction of the entire state.

Our approach can be physically motivated by assuming we have access to the experimental process where tomographic data is collected. If, at this stage, we were aware that we were restricted to an $m$-qubit reconstruction technique we could physically augment the $n$ qubit target state $\rho_{n}$ with an arbitrary system of $m-n$ single qubit states $\sigma$ to create the state
\begin{equation}
\label{eq:extension}
    \rho_{m}=\rho_{n}\otimes\sigma_{n+1}\otimes\sigma_{n+2}\otimes...\otimes\sigma_{m-n}.
\end{equation}
We could then collect standard tomographic measurement results for the total $\rho_{m}$ system, perform reconstruction and obtain $\tilde{\rho}_{m}$, the reconstruction of $\rho_{m}$.
Then, to obtain the desired result, the reconstruction of the $\rho_{n}$ state, we perform a partial trace $\tilde{\rho}_{n}=\text{Tr}_{\sigma}(\tilde{\rho}_{m})$ where $\text{Tr}_{\sigma}$ indicates tracing over the added single qubits.

However, not only is physically augmenting a quantum system experimentally challenging, but it is also unnecessary.
As pictured in Fig. \ref{fig:intro}, tomographic data for $\rho_{n}$ alone can be augmented with synthetic data for the added qubits, meaning that all modifications required to use an $m$-qubit reconstruction technique on an $n<m$ qubit system can be performed in postprocessing.
While in principle we could create synthetic measurement results that place the added single-qubit states $\sigma_{k}$ in any arbitrary state, it is conceptually simple to make them all completely mixed.

\begin{figure}[htb]
    \centering
    \includegraphics[width=\linewidth]{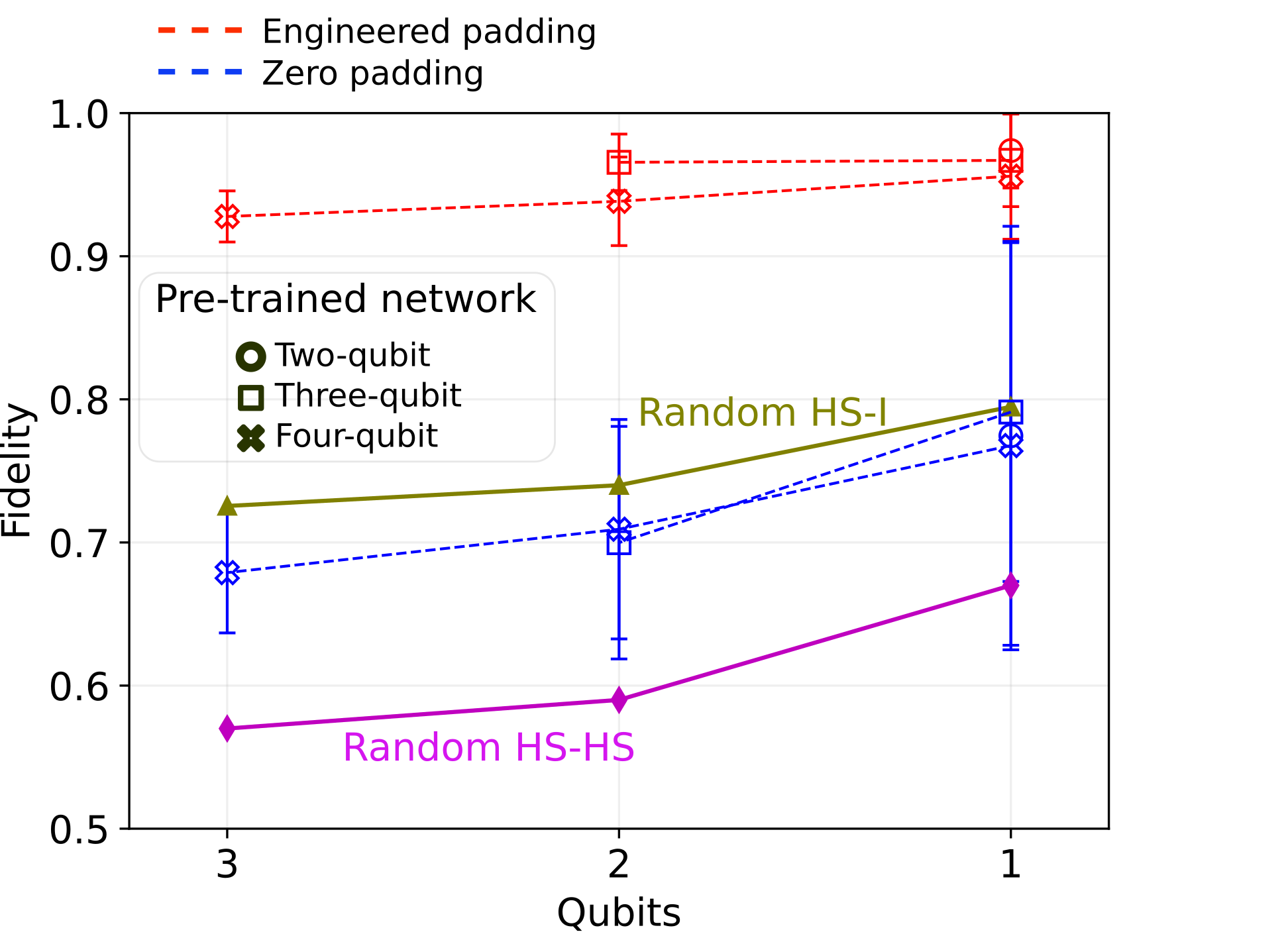}
    \caption{Reconstruction fidelity vs number of qubits. The markers $\times$, $\square$, and $\circ$, respectively, represent the reconstructing cases with the network pre-trained on $m=4$, $m=3$, and $m=2$ qubits. Similarly, the reconstruction fidelity with engineered padding and zero padding are respectively shown by red and blue dotted lines. The magenta line shows the average fidelity between two random density matrices sampled from the Hilbert-Schmidt (HS) measure, whereas the olive line represents the average fidelity between a maximally mixed state and a random density matrix sampled from the HS measure for a system of $n$ qubits.}
    \label{fig:mixed_vs_zero}
\end{figure}

The benefit of using separable and completely mixed single-qubit states is twofold. First, with each additional state being separable the joint measurement results are classical products of individual measurement results. Further, the completely mixed nature of the added states means that measurement outcomes are equal and exactly $1/2$ for all projective measurements in all orientations.
Therefore, postprocessing tomography data for $\rho_{n}$ to $\rho_{m}$ is merely a matter of multiplying each of the original measurement results by $(1/2)^{m-n}$ for a standard over-complete basis consisting of projections on each Pauli eigenvector.
The results of following this procedure using networks with $m\in\{2,3,4\}$ to reconstruct states with $n<m$ are shown in Fig. \ref{fig:mixed_vs_zero}. 

Given the expressive power of machine-learning-based reconstruction techniques it is reasonable to question if it is even necessary to perform the synthetic basis extension in order perform reconstruction with lower dimensional states.
For example, a naive alternative would be to merely zero pad all missing measurement results and use this as the input to the network, especially as the network is itself constrained to always produce a physical state.
More specifically, given an $m$-qubit network designed to take $6^m$ measurement outcomes as input, we could only fill the first $6^n$ measurement results zeroing out the remaining $6^m-6^n$ elements. This approach is not well motivated physically, as measurement inputs for the $m$-qubit system are joint measurements that include qubits unavailable to the $n$-qubit system. However, zero padding is nevertheless a straightforward way to perform $n$-qubit reconstruction with a network trained on $m$ qubits and is surprisingly effective when used to replace missing measurements for an $n=m$ qubit reconstruction~\cite{lohani2020machine}. Note that the order of the $6^n$ measurement results is such that the local bases correctly match their portion of the $m$-qubit joint measurements. 

To demonstrate the dramatic difference between this naive zero-padding approach and the basis augmentation approach above we have included the blue lines in Fig.~\ref{fig:mixed_vs_zero}.
These reconstructions were performed using the same networks as the red lines but with missing measurements completed by zero padding.
The separation in average fidelity between these two approaches is significant.
We compare the zero-padding approach to the trivial strategies of (i)~selecting another $n$-qubit state at random from the same distribution or (ii)~selecting the maximally mixed state always. Interestingly, we find that zero padding only performs marginally better than randomly selecting another state according to the HS measure and performs worse than always selecting identity. These results are shown visually in Fig.~\ref{fig:mixed_vs_zero} where the magenta line shows the average fidelity of two states chosen at random from the HS distribution, and the olive line shows when one is always the identity. More information is available in Appendix~\ref{app:HS}.

In light of the similar trends in Figs.~\ref{fig:fids_sub_system_all_HS} and \ref{fig:mixed_vs_zero}, it is certainly tempting to take the average fidelities found for $n=m$ as lower bounds for the $n<m$ cases. Yet whereas monotonicity justifies such a bound in Sec.~\ref{sec:reduced}, it does not apply to the results in Fig.~\ref{fig:mixed_vs_zero}.
This can be understood through examination of Eq.~\eqref{eq:extension}. Even if the $n$-qubit target states $\rho_n$ are HS-distributed in $2^n$ dimensions, the $m$-qubit extension $\rho_m$ is not HS-distributed in $2^m$: $m-n$ of its qubits are restricted to fixed states. Thus, any training results obtained on $m$-qubit HS-distributed quantum states cannot be used to bound the fidelities of $\rho_m$ defined in Eq.~\eqref{eq:extension}, which would have been required to thereafter bound the traced-down versions $\rho_n$. Nevertheless, despite the formal inapplicability of monotonicity, the observed scaling does match our initial intuition motivated by it: a single neural network is able to infer quantum states from Hilbert spaces of lower dimension than that on which it is trained, with fidelity even higher than the designed high-dimensional case.

\section{Conclusion}

In this work, we have proposed a physically motivated approach to performing quantum state reconstruction on systems of $n$ qubits when restricted to a state reconstruction technique intended for $m\geq n$ qubits. The utility of this approach is based on previous results indicating an approximately exponential separation in the required resources for training a neural network to perform quantum state reconstruction and reconstruction itself. Hence, efforts to avoid training an individual network for every potential system dimension that may be encountered in experimental scenarios can potentially offer significant resource savings. 

We began by describing a close link between the average reconstruction fidelity of $m$-qubit states and their $n<m$ qubit reduced density matrices using the well-known monotonicity bound of the fidelity. In particular, the average reconstruction fidelity of an $m$-qubit QST approach found during training can serve as the lower bound on the average reconstruction fidelity of any reduced density matrix from such an $m$-qubit reconstruction. 
As a proof of principle, we included an illustrative example based on simulated quantum state tomography measurements using a pre-trained machine-learning-based state reconstruction system for $m\in\{2,3,4\}$. 
We performed reconstruction for all $m$-qubit systems for each of the pre-trained networks and compared their average $m$-qubit performance to the fidelity of their reduced density matrices. We found the average reconstruction fidelities outperformed the lower bound due to the monotonicity in all cases.

Finally, given the high-performance average reconstruction of reduced density matrices, both implied by the monotonicity bound and confirmed in our numerical results, we proposed a method for leveraging $m$-qubit QST systems to perform $n<m$ qubit reconstructions.
Our approach consists of expanding $n$ qubits to $m>n$ qubit systems via postprocessing and then recovering the $n$-qubit density matrix through partial trace. In particular, we propose augmenting the collected tomography data with results from fictitious single-qubit states.
In our study, we have opted to use completely mixed single-qubit states to achieve this due to their isotropic behavior under projective measurement and the relative simplicity of how these states alter joint measurements, i.e., as a multiplicative factor.
We demonstrate the proof of principle of this approach using systems of up to four qubits. 
Further, we compare the performance of our technique with the naive approach of expanding the dimensions through zero padding. The zero-padding method performs significantly worse than our simulated measurement approach and only marginally better than the theoretical lower bound of sidestepping tomography and randomly guessing an answer.

While we limited our discussion to systems based on qubits and collections of qubits, restricting the possible Hilbert space dimensions to powers of two, extensions to arbitrary dimensions are straightforward. Further, based on previous results showing the impact of engineering training sets to emphasize specific system features \cite{lohani2021improving,lohani2022data}, further improvements could potentially be found by developing training sets that explicitly consider the distribution of their reduced density matrices.

\begin{acknowledgments}
Work by S. Lohani and T. A. Searles was supported in part by the U.S. Department of Energy, Office of Science, National Quantum Information Science Research Centers, Co-design Center for Quantum Advantage (C2QA) under contract number DE-SC0012704. A portion of this work was performed at Oak Ridge National Laboratory, operated by UT-Battelle for the U.S. Department of Energy under contract no. DE-AC05-00OR22725. J. M. Lukens acknowledges funding by the U.S. Department of Energy, Office of Science, Advanced Scientific Computing Research, through the Early Career Research Program (Field Work Proposal ERKJ353).  The views and conclusions contained in this document are those of the authors and should not be interpreted as representing the official policies, either expressed or implied, of the Army Research Laboratory or the U.S. Government. The U.S. Government is authorized to reproduce and distribute reprints for Government purposes notwithstanding any copyright notation herein. Additionally, this material is based upon work supported by, or in part by, the Army Research Laboratory and the Army Research Office under contract/grant numbers W911NF-19-2-0087 and W911NF-20-2-0168.
\end{acknowledgments}



\appendix

\section{Results when training and testing states are sampled according to the Bures distribution}
\label{APP:Bures}
\begin{figure}[htb]
    \centering
    \includegraphics[width=\linewidth]{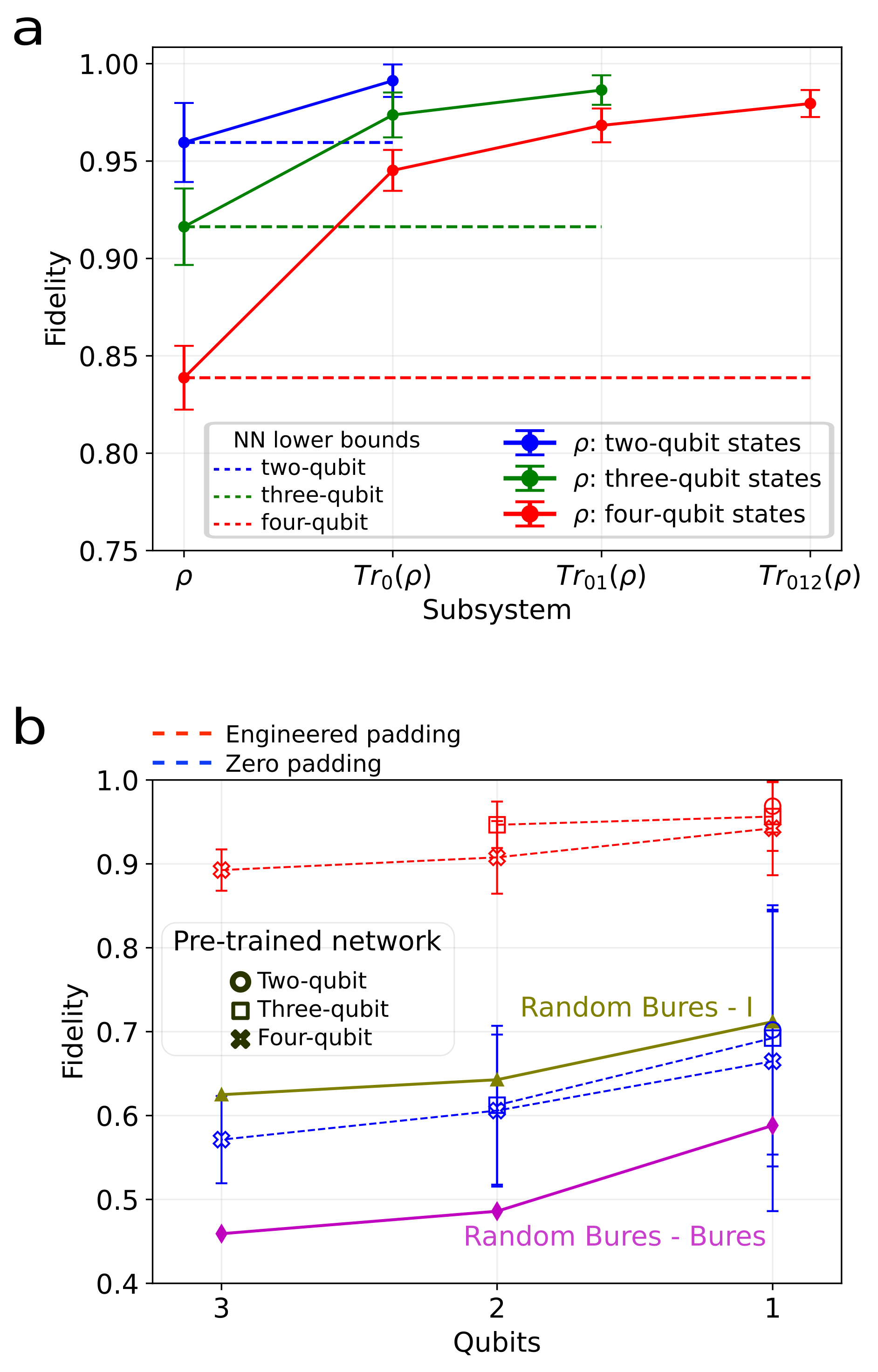}
    \caption{Test and train with random quantum states sampled from the Bures metric. (a) Reconstruction fidelity with respect to subsystem of predicted quantum states. (b) Reconstruction fidelity versus number of qubits.}
    \label{fig:bures_appendix}
\end{figure}

In order to illustrate the concept for other cases, we sample 35,500 random quantum states $\rho$ according to the Bures measure for the given $m$~\cite{al2010random}. Similarly, we also simulate the associated $6^m$ Pauli measurement outcomes for systems with $m\in\{2,3,4\}$ qubits directly from expectation values. As described in the main text, we split the sampled data into a training set of size 35,000 and a validation set of size 500 to cross-validate the network performance per epoch. We implement a batch size of 100 in the training of a network. After training, we generate test sets, again, using the Bures measure for the same and lower qubit systems that are entirely unknown to the trained network. Finally, the reconstruction fidelity with respect to subsystem size and number of qubits are, respectively, shown in Fig.~\ref{fig:bures_appendix}(a) and (b). 
Although the average reconstruction fidelities for states sampled according to the Bures metric are slightly lower than the those drawn from the HS metric [see Fig.~\ref{fig:fids_sub_system_all_HS} in the main text], the same important scaling trends hold.

\section{Average fidelity between random quantum states}
\label{app:HS}



For completeness we include here the expression for the average fidelity $\langle F\rangle_{N}$ between two random mixed states of dimension $N$ generated according to the HS measure.
We take our results from \cite{zyczkowski2005average} where a more general expression applicable to two random mixed states chosen according to an arbitrary induced measure is presented.
In simplifying the results of \cite{zyczkowski2005average} we find
\begin{equation}
\begin{aligned}
    \langle F\rangle_{N}&=\frac{1}{N^4}\left[\text{Tr}\left(X_{0}^{-1}X_{1}\right) \right.\\
    &+\left.\left(\text{Tr}\left[X_{0}^{-1}X_{1/2}\right] \right)^{2}-\text{Tr}\left(\left[X_{0}^{-1}X_{1/2}\right]^{2}\right) \right]\\
    \end{aligned}
\end{equation}
where $X_{n}$ defines a matrix with entries
\begin{equation}
    (X_{n})_{k,l}=\Gamma\left(n+k+l-1\right)\Gamma(n+1),
\end{equation}
for $k,l\in\{1,2,...,N\}$, and $\Gamma(\cdot)$ is the usual gamma function.
Using these expressions we find for one, two, and three qubit states, respectively, that $\langle F\rangle_{2}=0.67$, $\langle F\rangle_{4}=0.59$, and $\langle F\rangle_{8}=0.57$.

In addition to the average fidelity between two random density matrices chosen according to the HS measure, we also show three other average fidelities in Figs.~\ref{fig:mixed_vs_zero} and \ref{fig:bures_appendix}(b). Figure~\ref{fig:bures_appendix}(b) includes the average fidelity between two random quantum states chosen according to the Bures measure. In \cite{zyczkowski2005average}, an analytical result is given for this situation in the case of single-qubit states: $\braket{F}_{2}=0.590$, which we supplement with numerical results for two and three qubits to create the relevant curve in Fig.~\ref{fig:bures_appendix}b. Finally, in Figs.~\ref{fig:mixed_vs_zero} and \ref{fig:bures_appendix}(b), we also show the average fidelity between random states chosen according to either the HS or Bures measures against the maximally mixed state. We obtain these values numerically but note that asymptotic results for these situations are available~\cite{zyczkowski2005average}.

\newpage
\bibliographystyle{apsrev4-1}
\bibliography{refs-QTML}

\begin{thebibliography}{55}%
\makeatletter
\providecommand \@ifxundefined [1]{%
 \@ifx{#1\undefined}
}%
\providecommand \@ifnum [1]{%
 \ifnum #1\expandafter \@firstoftwo
 \else \expandafter \@secondoftwo
 \fi
}%
\providecommand \@ifx [1]{%
 \ifx #1\expandafter \@firstoftwo
 \else \expandafter \@secondoftwo
 \fi
}%
\providecommand \natexlab [1]{#1}%
\providecommand \enquote  [1]{``#1''}%
\providecommand \bibnamefont  [1]{#1}%
\providecommand \bibfnamefont [1]{#1}%
\providecommand \citenamefont [1]{#1}%
\providecommand \href@noop [0]{\@secondoftwo}%
\providecommand \href [0]{\begingroup \@sanitize@url \@href}%
\providecommand \@href[1]{\@@startlink{#1}\@@href}%
\providecommand \@@href[1]{\endgroup#1\@@endlink}%
\providecommand \@sanitize@url [0]{\catcode `\\12\catcode `\$12\catcode
  `\&12\catcode `\#12\catcode `\^12\catcode `\_12\catcode `\%12\relax}%
\providecommand \@@startlink[1]{}%
\providecommand \@@endlink[0]{}%
\providecommand \url  [0]{\begingroup\@sanitize@url \@url }%
\providecommand \@url [1]{\endgroup\@href {#1}{\urlprefix }}%
\providecommand \urlprefix  [0]{URL }%
\providecommand \Eprint [0]{\href }%
\providecommand \doibase [0]{http://dx.doi.org/}%
\providecommand \selectlanguage [0]{\@gobble}%
\providecommand \bibinfo  [0]{\@secondoftwo}%
\providecommand \bibfield  [0]{\@secondoftwo}%
\providecommand \translation [1]{[#1]}%
\providecommand \BibitemOpen [0]{}%
\providecommand \bibitemStop [0]{}%
\providecommand \bibitemNoStop [0]{.\EOS\space}%
\providecommand \EOS [0]{\spacefactor3000\relax}%
\providecommand \BibitemShut  [1]{\csname bibitem#1\endcsname}%
\let\auto@bib@innerbib\@empty
\bibitem [{\citenamefont {Ekert}\ \emph {et~al.}(2002)\citenamefont {Ekert},
  \citenamefont {Alves}, \citenamefont {Oi}, \citenamefont {Horodecki},
  \citenamefont {Horodecki},\ and\ \citenamefont {Kwek}}]{ekert2002direct}%
  \BibitemOpen
  \bibfield  {author} {\bibinfo {author} {\bibfnamefont {A.~K.}\ \bibnamefont
  {Ekert}}, \bibinfo {author} {\bibfnamefont {C.~M.}\ \bibnamefont {Alves}},
  \bibinfo {author} {\bibfnamefont {D.~K.}\ \bibnamefont {Oi}}, \bibinfo
  {author} {\bibfnamefont {M.}~\bibnamefont {Horodecki}}, \bibinfo {author}
  {\bibfnamefont {P.}~\bibnamefont {Horodecki}}, \ and\ \bibinfo {author}
  {\bibfnamefont {L.~C.}\ \bibnamefont {Kwek}},\ }\href@noop {} {\bibfield
  {journal} {\bibinfo  {journal} {Phys. Rev. Lett.}\ }\textbf {\bibinfo
  {volume} {88}},\ \bibinfo {pages} {217901} (\bibinfo {year}
  {2002})}\BibitemShut {NoStop}%
\bibitem [{\citenamefont {Flammia}\ and\ \citenamefont
  {Liu}(2011)}]{Flammia2011}%
  \BibitemOpen
  \bibfield  {author} {\bibinfo {author} {\bibfnamefont {S.~T.}\ \bibnamefont
  {Flammia}}\ and\ \bibinfo {author} {\bibfnamefont {Y.-K.}\ \bibnamefont
  {Liu}},\ }\href {\doibase 10.1103/PhysRevLett.106.230501} {\bibfield
  {journal} {\bibinfo  {journal} {Phys. Rev. Lett.}\ }\textbf {\bibinfo
  {volume} {106}},\ \bibinfo {pages} {230501} (\bibinfo {year}
  {2011})}\BibitemShut {NoStop}%
\bibitem [{\citenamefont {Spengler}\ \emph {et~al.}(2012)\citenamefont
  {Spengler}, \citenamefont {Huber}, \citenamefont {Brierley}, \citenamefont
  {Adaktylos},\ and\ \citenamefont {Hiesmayr}}]{Spengler2012}%
  \BibitemOpen
  \bibfield  {author} {\bibinfo {author} {\bibfnamefont {C.}~\bibnamefont
  {Spengler}}, \bibinfo {author} {\bibfnamefont {M.}~\bibnamefont {Huber}},
  \bibinfo {author} {\bibfnamefont {S.}~\bibnamefont {Brierley}}, \bibinfo
  {author} {\bibfnamefont {T.}~\bibnamefont {Adaktylos}}, \ and\ \bibinfo
  {author} {\bibfnamefont {B.~C.}\ \bibnamefont {Hiesmayr}},\ }\href {\doibase
  10.1103/PhysRevA.86.022311} {\bibfield  {journal} {\bibinfo  {journal} {Phys.
  Rev. A}\ }\textbf {\bibinfo {volume} {86}},\ \bibinfo {pages} {022311}
  (\bibinfo {year} {2012})}\BibitemShut {NoStop}%
\bibitem [{\citenamefont {Huang}\ \emph {et~al.}(2020)\citenamefont {Huang},
  \citenamefont {Kueng},\ and\ \citenamefont {Preskill}}]{huang2020predicting}%
  \BibitemOpen
  \bibfield  {author} {\bibinfo {author} {\bibfnamefont {H.-Y.}\ \bibnamefont
  {Huang}}, \bibinfo {author} {\bibfnamefont {R.}~\bibnamefont {Kueng}}, \ and\
  \bibinfo {author} {\bibfnamefont {J.}~\bibnamefont {Preskill}},\ }\href@noop
  {} {\bibfield  {journal} {\bibinfo  {journal} {Nat. Phys.}\ }\textbf
  {\bibinfo {volume} {16}},\ \bibinfo {pages} {1050} (\bibinfo {year}
  {2020})}\BibitemShut {NoStop}%
\bibitem [{\citenamefont {Eisert}\ \emph {et~al.}(2020)\citenamefont {Eisert},
  \citenamefont {Hangleiter}, \citenamefont {Walk}, \citenamefont {Roth},
  \citenamefont {Markham}, \citenamefont {Parekh}, \citenamefont {Chabaud},\
  and\ \citenamefont {Kashefi}}]{Eisert2020}%
  \BibitemOpen
  \bibfield  {author} {\bibinfo {author} {\bibfnamefont {J.}~\bibnamefont
  {Eisert}}, \bibinfo {author} {\bibfnamefont {D.}~\bibnamefont {Hangleiter}},
  \bibinfo {author} {\bibfnamefont {N.}~\bibnamefont {Walk}}, \bibinfo {author}
  {\bibfnamefont {I.}~\bibnamefont {Roth}}, \bibinfo {author} {\bibfnamefont
  {D.}~\bibnamefont {Markham}}, \bibinfo {author} {\bibfnamefont
  {R.}~\bibnamefont {Parekh}}, \bibinfo {author} {\bibfnamefont
  {U.}~\bibnamefont {Chabaud}}, \ and\ \bibinfo {author} {\bibfnamefont
  {E.}~\bibnamefont {Kashefi}},\ }\href@noop {} {\bibfield  {journal} {\bibinfo
   {journal} {Nat. Rev. Phys.}\ }\textbf {\bibinfo {volume} {2}},\ \bibinfo
  {pages} {382} (\bibinfo {year} {2020})}\BibitemShut {NoStop}%
\bibitem [{\citenamefont {Lukens}\ \emph {et~al.}(2021)\citenamefont {Lukens},
  \citenamefont {Law},\ and\ \citenamefont {Bennink}}]{lukens2021bayesian}%
  \BibitemOpen
  \bibfield  {author} {\bibinfo {author} {\bibfnamefont {J.~M.}\ \bibnamefont
  {Lukens}}, \bibinfo {author} {\bibfnamefont {K.~J.}\ \bibnamefont {Law}}, \
  and\ \bibinfo {author} {\bibfnamefont {R.~S.}\ \bibnamefont {Bennink}},\
  }\href@noop {} {\bibfield  {journal} {\bibinfo  {journal} {npj Quantum Inf.}\
  }\textbf {\bibinfo {volume} {7}},\ \bibinfo {pages} {113} (\bibinfo {year}
  {2021})}\BibitemShut {NoStop}%
\bibitem [{\citenamefont {Thew}\ \emph {et~al.}(2002)\citenamefont {Thew},
  \citenamefont {Nemoto}, \citenamefont {White},\ and\ \citenamefont
  {Munro}}]{thew2002qudit}%
  \BibitemOpen
  \bibfield  {author} {\bibinfo {author} {\bibfnamefont {R.~T.}\ \bibnamefont
  {Thew}}, \bibinfo {author} {\bibfnamefont {K.}~\bibnamefont {Nemoto}},
  \bibinfo {author} {\bibfnamefont {A.~G.}\ \bibnamefont {White}}, \ and\
  \bibinfo {author} {\bibfnamefont {W.~J.}\ \bibnamefont {Munro}},\ }\href@noop
  {} {\bibfield  {journal} {\bibinfo  {journal} {Phys. Rev. A}\ }\textbf
  {\bibinfo {volume} {66}},\ \bibinfo {pages} {012303} (\bibinfo {year}
  {2002})}\BibitemShut {NoStop}%
\bibitem [{\citenamefont {Altepeter}\ \emph {et~al.}(2005)\citenamefont
  {Altepeter}, \citenamefont {Jeffrey},\ and\ \citenamefont
  {Kwiat}}]{altepeter2005photonic}%
  \BibitemOpen
  \bibfield  {author} {\bibinfo {author} {\bibfnamefont {J.~B.}\ \bibnamefont
  {Altepeter}}, \bibinfo {author} {\bibfnamefont {E.~R.}\ \bibnamefont
  {Jeffrey}}, \ and\ \bibinfo {author} {\bibfnamefont {P.~G.}\ \bibnamefont
  {Kwiat}},\ }\href@noop {} {\bibfield  {journal} {\bibinfo  {journal} {Adv.
  At. Mol. Opt. Phys.}\ }\textbf {\bibinfo {volume} {52}},\ \bibinfo {pages}
  {105} (\bibinfo {year} {2005})}\BibitemShut {NoStop}%
\bibitem [{\citenamefont {James}\ \emph {et~al.}(2005)\citenamefont {James},
  \citenamefont {Kwiat}, \citenamefont {Munro},\ and\ \citenamefont
  {White}}]{james2005measurement}%
  \BibitemOpen
  \bibfield  {author} {\bibinfo {author} {\bibfnamefont {D.~F.}\ \bibnamefont
  {James}}, \bibinfo {author} {\bibfnamefont {P.~G.}\ \bibnamefont {Kwiat}},
  \bibinfo {author} {\bibfnamefont {W.~J.}\ \bibnamefont {Munro}}, \ and\
  \bibinfo {author} {\bibfnamefont {A.~G.}\ \bibnamefont {White}},\ }in\
  \href@noop {} {\emph {\bibinfo {booktitle} {Asymptotic Theory of Quantum
  Statistical Inference: Selected Papers}}}\ (\bibinfo  {publisher} {World
  Scientific},\ \bibinfo {year} {2005})\ pp.\ \bibinfo {pages}
  {509--538}\BibitemShut {NoStop}%
\bibitem [{\citenamefont {Hradil}(1997)}]{Hradil1997}%
  \BibitemOpen
  \bibfield  {author} {\bibinfo {author} {\bibfnamefont {Z.}~\bibnamefont
  {Hradil}},\ }\href {\doibase 10.1103/PhysRevA.55.R1561} {\bibfield  {journal}
  {\bibinfo  {journal} {Phys. Rev. A}\ }\textbf {\bibinfo {volume} {55}},\
  \bibinfo {pages} {R1561} (\bibinfo {year} {1997})}\BibitemShut {NoStop}%
\bibitem [{\citenamefont {Banaszek}\ \emph {et~al.}(1999)\citenamefont
  {Banaszek}, \citenamefont {D'Ariano}, \citenamefont {Paris},\ and\
  \citenamefont {Sacchi}}]{Banaszek1999}%
  \BibitemOpen
  \bibfield  {author} {\bibinfo {author} {\bibfnamefont {K.}~\bibnamefont
  {Banaszek}}, \bibinfo {author} {\bibfnamefont {G.~M.}\ \bibnamefont
  {D'Ariano}}, \bibinfo {author} {\bibfnamefont {M.~G.~A.}\ \bibnamefont
  {Paris}}, \ and\ \bibinfo {author} {\bibfnamefont {M.~F.}\ \bibnamefont
  {Sacchi}},\ }\href {\doibase 10.1103/PhysRevA.61.010304} {\bibfield
  {journal} {\bibinfo  {journal} {Phys. Rev. A}\ }\textbf {\bibinfo {volume}
  {61}},\ \bibinfo {pages} {010304} (\bibinfo {year} {1999})}\BibitemShut
  {NoStop}%
\bibitem [{\citenamefont {James}\ \emph {et~al.}(2001)\citenamefont {James},
  \citenamefont {Kwiat}, \citenamefont {Munro},\ and\ \citenamefont
  {White}}]{James2001}%
  \BibitemOpen
  \bibfield  {author} {\bibinfo {author} {\bibfnamefont {D.~F.~V.}\
  \bibnamefont {James}}, \bibinfo {author} {\bibfnamefont {P.~G.}\ \bibnamefont
  {Kwiat}}, \bibinfo {author} {\bibfnamefont {W.~J.}\ \bibnamefont {Munro}}, \
  and\ \bibinfo {author} {\bibfnamefont {A.~G.}\ \bibnamefont {White}},\ }\href
  {\doibase 10.1103/PhysRevA.64.052312} {\bibfield  {journal} {\bibinfo
  {journal} {Phys. Rev. A}\ }\textbf {\bibinfo {volume} {64}},\ \bibinfo
  {pages} {052312} (\bibinfo {year} {2001})}\BibitemShut {NoStop}%
\bibitem [{\citenamefont {Lvovsky}(2004)}]{Lvovsky2004}%
  \BibitemOpen
  \bibfield  {author} {\bibinfo {author} {\bibfnamefont {A.~I.}\ \bibnamefont
  {Lvovsky}},\ }\href {\doibase 10.1088/1464-4266/6/6/014} {\bibfield
  {journal} {\bibinfo  {journal} {J. Opt. B: Quantum Semiclass. Opt.}\ }\textbf
  {\bibinfo {volume} {6}},\ \bibinfo {pages} {S556} (\bibinfo {year}
  {2004})}\BibitemShut {NoStop}%
\bibitem [{\citenamefont {Smolin}\ \emph {et~al.}(2012)\citenamefont {Smolin},
  \citenamefont {Gambetta},\ and\ \citenamefont
  {Smith}}]{smolin_efficient_2012}%
  \BibitemOpen
  \bibfield  {author} {\bibinfo {author} {\bibfnamefont {J.~A.}\ \bibnamefont
  {Smolin}}, \bibinfo {author} {\bibfnamefont {J.~M.}\ \bibnamefont
  {Gambetta}}, \ and\ \bibinfo {author} {\bibfnamefont {G.}~\bibnamefont
  {Smith}},\ }\href {\doibase 10.1103/PhysRevLett.108.070502} {\bibfield
  {journal} {\bibinfo  {journal} {Phys. Rev. Lett.}\ }\textbf {\bibinfo
  {volume} {108}},\ \bibinfo {pages} {070502} (\bibinfo {year}
  {2012})}\BibitemShut {NoStop}%
\bibitem [{\citenamefont {Blume-Kohout}(2010)}]{Blume2010}%
  \BibitemOpen
  \bibfield  {author} {\bibinfo {author} {\bibfnamefont {R.}~\bibnamefont
  {Blume-Kohout}},\ }\href {http://stacks.iop.org/1367-2630/12/i=4/a=043034}
  {\bibfield  {journal} {\bibinfo  {journal} {New J. Phys.}\ }\textbf {\bibinfo
  {volume} {12}},\ \bibinfo {pages} {043034} (\bibinfo {year}
  {2010})}\BibitemShut {NoStop}%
\bibitem [{\citenamefont {Husz\'ar}\ and\ \citenamefont
  {Houlsby}(2012)}]{Huszar2012}%
  \BibitemOpen
  \bibfield  {author} {\bibinfo {author} {\bibfnamefont {F.}~\bibnamefont
  {Husz\'ar}}\ and\ \bibinfo {author} {\bibfnamefont {N.~M.~T.}\ \bibnamefont
  {Houlsby}},\ }\href {\doibase 10.1103/PhysRevA.85.052120} {\bibfield
  {journal} {\bibinfo  {journal} {Phys. Rev. A}\ }\textbf {\bibinfo {volume}
  {85}},\ \bibinfo {pages} {052120} (\bibinfo {year} {2012})}\BibitemShut
  {NoStop}%
\bibitem [{\citenamefont {Kravtsov}\ \emph {et~al.}(2013)\citenamefont
  {Kravtsov}, \citenamefont {Straupe}, \citenamefont {Radchenko}, \citenamefont
  {Houlsby}, \citenamefont {Husz\'ar},\ and\ \citenamefont
  {Kulik}}]{Kravtsov2013}%
  \BibitemOpen
  \bibfield  {author} {\bibinfo {author} {\bibfnamefont {K.~S.}\ \bibnamefont
  {Kravtsov}}, \bibinfo {author} {\bibfnamefont {S.~S.}\ \bibnamefont
  {Straupe}}, \bibinfo {author} {\bibfnamefont {I.~V.}\ \bibnamefont
  {Radchenko}}, \bibinfo {author} {\bibfnamefont {N.~M.~T.}\ \bibnamefont
  {Houlsby}}, \bibinfo {author} {\bibfnamefont {F.}~\bibnamefont {Husz\'ar}}, \
  and\ \bibinfo {author} {\bibfnamefont {S.~P.}\ \bibnamefont {Kulik}},\ }\href
  {\doibase 10.1103/PhysRevA.87.062122} {\bibfield  {journal} {\bibinfo
  {journal} {Phys. Rev. A}\ }\textbf {\bibinfo {volume} {87}},\ \bibinfo
  {pages} {062122} (\bibinfo {year} {2013})}\BibitemShut {NoStop}%
\bibitem [{\citenamefont {Seah}\ \emph {et~al.}(2015)\citenamefont {Seah},
  \citenamefont {Shang}, \citenamefont {Ng}, \citenamefont {Nott},\ and\
  \citenamefont {Englert}}]{Seah2015}%
  \BibitemOpen
  \bibfield  {author} {\bibinfo {author} {\bibfnamefont {Y.-L.}\ \bibnamefont
  {Seah}}, \bibinfo {author} {\bibfnamefont {J.}~\bibnamefont {Shang}},
  \bibinfo {author} {\bibfnamefont {H.~K.}\ \bibnamefont {Ng}}, \bibinfo
  {author} {\bibfnamefont {D.~J.}\ \bibnamefont {Nott}}, \ and\ \bibinfo
  {author} {\bibfnamefont {B.-G.}\ \bibnamefont {Englert}},\ }\href {\doibase
  10.1088/1367-2630/17/4/043018} {\bibfield  {journal} {\bibinfo  {journal}
  {New J. Phys.}\ }\textbf {\bibinfo {volume} {17}},\ \bibinfo {pages} {043018}
  (\bibinfo {year} {2015})}\BibitemShut {NoStop}%
\bibitem [{\citenamefont {Granade}\ \emph {et~al.}(2016)\citenamefont
  {Granade}, \citenamefont {Combes},\ and\ \citenamefont {Cory}}]{Granade2016}%
  \BibitemOpen
  \bibfield  {author} {\bibinfo {author} {\bibfnamefont {C.}~\bibnamefont
  {Granade}}, \bibinfo {author} {\bibfnamefont {J.}~\bibnamefont {Combes}}, \
  and\ \bibinfo {author} {\bibfnamefont {D.~G.}\ \bibnamefont {Cory}},\ }\href
  {http://stacks.iop.org/1367-2630/18/i=3/a=033024} {\bibfield  {journal}
  {\bibinfo  {journal} {New J. Phys.}\ }\textbf {\bibinfo {volume} {18}},\
  \bibinfo {pages} {033024} (\bibinfo {year} {2016})}\BibitemShut {NoStop}%
\bibitem [{\citenamefont {Williams}\ and\ \citenamefont
  {Lougovski}(2017)}]{Williams2017}%
  \BibitemOpen
  \bibfield  {author} {\bibinfo {author} {\bibfnamefont {B.~P.}\ \bibnamefont
  {Williams}}\ and\ \bibinfo {author} {\bibfnamefont {P.}~\bibnamefont
  {Lougovski}},\ }\href {http://stacks.iop.org/1367-2630/19/i=4/a=043003}
  {\bibfield  {journal} {\bibinfo  {journal} {New J. Phys.}\ }\textbf {\bibinfo
  {volume} {19}},\ \bibinfo {pages} {043003} (\bibinfo {year}
  {2017})}\BibitemShut {NoStop}%
\bibitem [{\citenamefont {Mai}\ and\ \citenamefont {Alquier}(2017)}]{Mai2017}%
  \BibitemOpen
  \bibfield  {author} {\bibinfo {author} {\bibfnamefont {T.~T.}\ \bibnamefont
  {Mai}}\ and\ \bibinfo {author} {\bibfnamefont {P.}~\bibnamefont {Alquier}},\
  }\href {\doibase https://doi.org/10.1016/j.jspi.2016.11.003} {\bibfield
  {journal} {\bibinfo  {journal} {J. Stat. Plan. Inference}\ }\textbf {\bibinfo
  {volume} {184}},\ \bibinfo {pages} {62} (\bibinfo {year} {2017})}\BibitemShut
  {NoStop}%
\bibitem [{\citenamefont {Lukens}\ \emph {et~al.}(2020)\citenamefont {Lukens},
  \citenamefont {Law}, \citenamefont {Jasra},\ and\ \citenamefont
  {Lougovski}}]{lukens2020practical}%
  \BibitemOpen
  \bibfield  {author} {\bibinfo {author} {\bibfnamefont {J.~M.}\ \bibnamefont
  {Lukens}}, \bibinfo {author} {\bibfnamefont {K.~J.~H.}\ \bibnamefont {Law}},
  \bibinfo {author} {\bibfnamefont {A.}~\bibnamefont {Jasra}}, \ and\ \bibinfo
  {author} {\bibfnamefont {P.}~\bibnamefont {Lougovski}},\ }\href {\doibase
  10.1088/1367-2630/ab8efa} {\bibfield  {journal} {\bibinfo  {journal} {New J.
  Phys.}\ }\textbf {\bibinfo {volume} {22}},\ \bibinfo {pages} {063038}
  (\bibinfo {year} {2020})}\BibitemShut {NoStop}%
\bibitem [{\citenamefont {Simmerman}\ \emph {et~al.}(2020)\citenamefont
  {Simmerman}, \citenamefont {Lu}, \citenamefont {Weiner},\ and\ \citenamefont
  {Lukens}}]{simmerman2020efficient}%
  \BibitemOpen
  \bibfield  {author} {\bibinfo {author} {\bibfnamefont {E.~M.}\ \bibnamefont
  {Simmerman}}, \bibinfo {author} {\bibfnamefont {H.-H.}\ \bibnamefont {Lu}},
  \bibinfo {author} {\bibfnamefont {A.~M.}\ \bibnamefont {Weiner}}, \ and\
  \bibinfo {author} {\bibfnamefont {J.~M.}\ \bibnamefont {Lukens}},\
  }\href@noop {} {\bibfield  {journal} {\bibinfo  {journal} {Opt. Lett.}\
  }\textbf {\bibinfo {volume} {45}},\ \bibinfo {pages} {2886} (\bibinfo {year}
  {2020})}\BibitemShut {NoStop}%
\bibitem [{\citenamefont {Lu}\ \emph {et~al.}(2021)\citenamefont {Lu},
  \citenamefont {Myilswamy}, \citenamefont {Bennink}, \citenamefont {Seshadri},
  \citenamefont {Alshaykh}, \citenamefont {Liu}, \citenamefont {Kippenberg},
  \citenamefont {Leaird}, \citenamefont {Weiner},\ and\ \citenamefont
  {Lukens}}]{Lu2021}%
  \BibitemOpen
  \bibfield  {author} {\bibinfo {author} {\bibfnamefont {H.-H.}\ \bibnamefont
  {Lu}}, \bibinfo {author} {\bibfnamefont {K.~V.}\ \bibnamefont {Myilswamy}},
  \bibinfo {author} {\bibfnamefont {R.~S.}\ \bibnamefont {Bennink}}, \bibinfo
  {author} {\bibfnamefont {S.}~\bibnamefont {Seshadri}}, \bibinfo {author}
  {\bibfnamefont {M.~S.}\ \bibnamefont {Alshaykh}}, \bibinfo {author}
  {\bibfnamefont {J.}~\bibnamefont {Liu}}, \bibinfo {author} {\bibfnamefont
  {T.~J.}\ \bibnamefont {Kippenberg}}, \bibinfo {author} {\bibfnamefont
  {D.~E.}\ \bibnamefont {Leaird}}, \bibinfo {author} {\bibfnamefont {A.~M.}\
  \bibnamefont {Weiner}}, \ and\ \bibinfo {author} {\bibfnamefont {J.~M.}\
  \bibnamefont {Lukens}},\ }\href@noop {} {\bibfield  {journal} {\bibinfo
  {journal} {arXiv:2108.04124}\ } (\bibinfo {year} {2021})}\BibitemShut
  {NoStop}%
\bibitem [{\citenamefont {Chapman}\ \emph {et~al.}(2022)\citenamefont
  {Chapman}, \citenamefont {Lukens}, \citenamefont {Qi}, \citenamefont
  {Pooser},\ and\ \citenamefont {Peters}}]{chapman2022}%
  \BibitemOpen
  \bibfield  {author} {\bibinfo {author} {\bibfnamefont {J.~C.}\ \bibnamefont
  {Chapman}}, \bibinfo {author} {\bibfnamefont {J.~M.}\ \bibnamefont {Lukens}},
  \bibinfo {author} {\bibfnamefont {B.}~\bibnamefont {Qi}}, \bibinfo {author}
  {\bibfnamefont {R.~C.}\ \bibnamefont {Pooser}}, \ and\ \bibinfo {author}
  {\bibfnamefont {N.~A.}\ \bibnamefont {Peters}},\ }\href {\doibase
  10.1364/OE.456597} {\bibfield  {journal} {\bibinfo  {journal} {Opt. Express}\
  }\textbf {\bibinfo {volume} {30}},\ \bibinfo {pages} {15184} (\bibinfo {year}
  {2022})}\BibitemShut {NoStop}%
\bibitem [{\citenamefont {Lu}\ \emph {et~al.}(2018)\citenamefont {Lu},
  \citenamefont {Huang}, \citenamefont {Li}, \citenamefont {Li}, \citenamefont
  {Chen}, \citenamefont {Lu}, \citenamefont {Ji}, \citenamefont {Shen},
  \citenamefont {Zhou},\ and\ \citenamefont {Zeng}}]{lu2018separability}%
  \BibitemOpen
  \bibfield  {author} {\bibinfo {author} {\bibfnamefont {S.}~\bibnamefont
  {Lu}}, \bibinfo {author} {\bibfnamefont {S.}~\bibnamefont {Huang}}, \bibinfo
  {author} {\bibfnamefont {K.}~\bibnamefont {Li}}, \bibinfo {author}
  {\bibfnamefont {J.}~\bibnamefont {Li}}, \bibinfo {author} {\bibfnamefont
  {J.}~\bibnamefont {Chen}}, \bibinfo {author} {\bibfnamefont {D.}~\bibnamefont
  {Lu}}, \bibinfo {author} {\bibfnamefont {Z.}~\bibnamefont {Ji}}, \bibinfo
  {author} {\bibfnamefont {Y.}~\bibnamefont {Shen}}, \bibinfo {author}
  {\bibfnamefont {D.}~\bibnamefont {Zhou}}, \ and\ \bibinfo {author}
  {\bibfnamefont {B.}~\bibnamefont {Zeng}},\ }\href@noop {} {\bibfield
  {journal} {\bibinfo  {journal} {Phys. Rev. A}\ }\textbf {\bibinfo {volume}
  {98}},\ \bibinfo {pages} {012315} (\bibinfo {year} {2018})}\BibitemShut
  {NoStop}%
\bibitem [{\citenamefont {Lohani}\ \emph
  {et~al.}(2020{\natexlab{a}})\citenamefont {Lohani}, \citenamefont {Kirby},
  \citenamefont {Brodsky}, \citenamefont {Danaci},\ and\ \citenamefont
  {Glasser}}]{lohani2020machine}%
  \BibitemOpen
  \bibfield  {author} {\bibinfo {author} {\bibfnamefont {S.}~\bibnamefont
  {Lohani}}, \bibinfo {author} {\bibfnamefont {B.~T.}\ \bibnamefont {Kirby}},
  \bibinfo {author} {\bibfnamefont {M.}~\bibnamefont {Brodsky}}, \bibinfo
  {author} {\bibfnamefont {O.}~\bibnamefont {Danaci}}, \ and\ \bibinfo {author}
  {\bibfnamefont {R.~T.}\ \bibnamefont {Glasser}},\ }\href {\doibase
  10.1088/2632-2153/ab9a21} {\bibfield  {journal} {\bibinfo  {journal} {Mach.
  Learn.: Sci. Technol.}\ }\textbf {\bibinfo {volume} {1}},\ \bibinfo {pages}
  {035007} (\bibinfo {year} {2020}{\natexlab{a}})}\BibitemShut {NoStop}%
\bibitem [{\citenamefont {Danaci}\ \emph {et~al.}(2021)\citenamefont {Danaci},
  \citenamefont {Lohani}, \citenamefont {Kirby},\ and\ \citenamefont
  {Glasser}}]{danaci2021machine}%
  \BibitemOpen
  \bibfield  {author} {\bibinfo {author} {\bibfnamefont {O.}~\bibnamefont
  {Danaci}}, \bibinfo {author} {\bibfnamefont {S.}~\bibnamefont {Lohani}},
  \bibinfo {author} {\bibfnamefont {B.~T.}\ \bibnamefont {Kirby}}, \ and\
  \bibinfo {author} {\bibfnamefont {R.~T.}\ \bibnamefont {Glasser}},\
  }\href@noop {} {\bibfield  {journal} {\bibinfo  {journal} {Machine Learning:
  Science and Technology}\ }\textbf {\bibinfo {volume} {2}},\ \bibinfo {pages}
  {035014} (\bibinfo {year} {2021})}\BibitemShut {NoStop}%
\bibitem [{\citenamefont {Ahmed}\ \emph
  {et~al.}(2021{\natexlab{a}})\citenamefont {Ahmed}, \citenamefont {Mu{\~n}oz},
  \citenamefont {Nori},\ and\ \citenamefont
  {Kockum}}]{ahmed2021classification}%
  \BibitemOpen
  \bibfield  {author} {\bibinfo {author} {\bibfnamefont {S.}~\bibnamefont
  {Ahmed}}, \bibinfo {author} {\bibfnamefont {C.~S.}\ \bibnamefont
  {Mu{\~n}oz}}, \bibinfo {author} {\bibfnamefont {F.}~\bibnamefont {Nori}}, \
  and\ \bibinfo {author} {\bibfnamefont {A.~F.}\ \bibnamefont {Kockum}},\
  }\href@noop {} {\bibfield  {journal} {\bibinfo  {journal} {Phys. Rev.
  Research}\ }\textbf {\bibinfo {volume} {3}},\ \bibinfo {pages} {033278}
  (\bibinfo {year} {2021}{\natexlab{a}})}\BibitemShut {NoStop}%
\bibitem [{\citenamefont {Lohani}\ \emph
  {et~al.}(2021{\natexlab{a}})\citenamefont {Lohani}, \citenamefont {Searles},
  \citenamefont {Kirby},\ and\ \citenamefont
  {Glasser}}]{lohani2021experimental}%
  \BibitemOpen
  \bibfield  {author} {\bibinfo {author} {\bibfnamefont {S.}~\bibnamefont
  {Lohani}}, \bibinfo {author} {\bibfnamefont {T.~A.}\ \bibnamefont {Searles}},
  \bibinfo {author} {\bibfnamefont {B.~T.}\ \bibnamefont {Kirby}}, \ and\
  \bibinfo {author} {\bibfnamefont {R.}~\bibnamefont {Glasser}},\ }\href@noop
  {} {\bibfield  {journal} {\bibinfo  {journal} {IEEE Trans. Quantum Eng.}\ ,\
  \bibinfo {pages} {2103410}} (\bibinfo {year}
  {2021}{\natexlab{a}})}\BibitemShut {NoStop}%
\bibitem [{\citenamefont {Lohani}\ \emph
  {et~al.}(2021{\natexlab{b}})\citenamefont {Lohani}, \citenamefont {Lukens},
  \citenamefont {Jones}, \citenamefont {Searles}, \citenamefont {Glasser},\
  and\ \citenamefont {Kirby}}]{lohani2021improving}%
  \BibitemOpen
  \bibfield  {author} {\bibinfo {author} {\bibfnamefont {S.}~\bibnamefont
  {Lohani}}, \bibinfo {author} {\bibfnamefont {J.~M.}\ \bibnamefont {Lukens}},
  \bibinfo {author} {\bibfnamefont {D.~E.}\ \bibnamefont {Jones}}, \bibinfo
  {author} {\bibfnamefont {T.~A.}\ \bibnamefont {Searles}}, \bibinfo {author}
  {\bibfnamefont {R.~T.}\ \bibnamefont {Glasser}}, \ and\ \bibinfo {author}
  {\bibfnamefont {B.~T.}\ \bibnamefont {Kirby}},\ }\href@noop {} {\bibfield
  {journal} {\bibinfo  {journal} {Phys. Rev. Research}\ }\textbf {\bibinfo
  {volume} {3}},\ \bibinfo {pages} {043145} (\bibinfo {year}
  {2021}{\natexlab{b}})}\BibitemShut {NoStop}%
\bibitem [{\citenamefont {Lohani}\ \emph {et~al.}(2022)\citenamefont {Lohani},
  \citenamefont {Lukens}, \citenamefont {Glasser}, \citenamefont {Searles},\
  and\ \citenamefont {Kirby}}]{lohani2022data}%
  \BibitemOpen
  \bibfield  {author} {\bibinfo {author} {\bibfnamefont {S.}~\bibnamefont
  {Lohani}}, \bibinfo {author} {\bibfnamefont {J.~M.}\ \bibnamefont {Lukens}},
  \bibinfo {author} {\bibfnamefont {R.~T.}\ \bibnamefont {Glasser}}, \bibinfo
  {author} {\bibfnamefont {T.~A.}\ \bibnamefont {Searles}}, \ and\ \bibinfo
  {author} {\bibfnamefont {B.~T.}\ \bibnamefont {Kirby}},\ }\href@noop {}
  {\bibfield  {journal} {\bibinfo  {journal} {arXiv:2201.09134}\ } (\bibinfo
  {year} {2022})}\BibitemShut {NoStop}%
\bibitem [{\citenamefont {Torlai}\ \emph {et~al.}(2018)\citenamefont {Torlai},
  \citenamefont {Mazzola}, \citenamefont {Carrasquilla}, \citenamefont
  {Troyer}, \citenamefont {Melko},\ and\ \citenamefont
  {Carleo}}]{torlai2018neural}%
  \BibitemOpen
  \bibfield  {author} {\bibinfo {author} {\bibfnamefont {G.}~\bibnamefont
  {Torlai}}, \bibinfo {author} {\bibfnamefont {G.}~\bibnamefont {Mazzola}},
  \bibinfo {author} {\bibfnamefont {J.}~\bibnamefont {Carrasquilla}}, \bibinfo
  {author} {\bibfnamefont {M.}~\bibnamefont {Troyer}}, \bibinfo {author}
  {\bibfnamefont {R.}~\bibnamefont {Melko}}, \ and\ \bibinfo {author}
  {\bibfnamefont {G.}~\bibnamefont {Carleo}},\ }\href@noop {} {\bibfield
  {journal} {\bibinfo  {journal} {Nat. Phys.}\ }\textbf {\bibinfo {volume}
  {14}},\ \bibinfo {pages} {447} (\bibinfo {year} {2018})}\BibitemShut
  {NoStop}%
\bibitem [{\citenamefont {Torlai}\ \emph {et~al.}(2019)\citenamefont {Torlai},
  \citenamefont {Timar}, \citenamefont {van Nieuwenburg}, \citenamefont
  {Levine}, \citenamefont {Omran}, \citenamefont {Keesling}, \citenamefont
  {Bernien}, \citenamefont {Greiner}, \citenamefont
  {Vuleti\ifmmode~\acute{c}\else \'{c}\fi{}}, \citenamefont {Lukin},
  \citenamefont {Melko},\ and\ \citenamefont {Endres}}]{torlai2019integrating}%
  \BibitemOpen
  \bibfield  {author} {\bibinfo {author} {\bibfnamefont {G.}~\bibnamefont
  {Torlai}}, \bibinfo {author} {\bibfnamefont {B.}~\bibnamefont {Timar}},
  \bibinfo {author} {\bibfnamefont {E.~P.~L.}\ \bibnamefont {van Nieuwenburg}},
  \bibinfo {author} {\bibfnamefont {H.}~\bibnamefont {Levine}}, \bibinfo
  {author} {\bibfnamefont {A.}~\bibnamefont {Omran}}, \bibinfo {author}
  {\bibfnamefont {A.}~\bibnamefont {Keesling}}, \bibinfo {author}
  {\bibfnamefont {H.}~\bibnamefont {Bernien}}, \bibinfo {author} {\bibfnamefont
  {M.}~\bibnamefont {Greiner}}, \bibinfo {author} {\bibfnamefont
  {V.}~\bibnamefont {Vuleti\ifmmode~\acute{c}\else \'{c}\fi{}}}, \bibinfo
  {author} {\bibfnamefont {M.~D.}\ \bibnamefont {Lukin}}, \bibinfo {author}
  {\bibfnamefont {R.~G.}\ \bibnamefont {Melko}}, \ and\ \bibinfo {author}
  {\bibfnamefont {M.}~\bibnamefont {Endres}},\ }\href {\doibase
  10.1103/PhysRevLett.123.230504} {\bibfield  {journal} {\bibinfo  {journal}
  {Phys. Rev. Lett.}\ }\textbf {\bibinfo {volume} {123}},\ \bibinfo {pages}
  {230504} (\bibinfo {year} {2019})}\BibitemShut {NoStop}%
\bibitem [{\citenamefont {Melkani}\ \emph {et~al.}(2020)\citenamefont
  {Melkani}, \citenamefont {Gneiting},\ and\ \citenamefont
  {Nori}}]{melkani2020eigenstate}%
  \BibitemOpen
  \bibfield  {author} {\bibinfo {author} {\bibfnamefont {A.}~\bibnamefont
  {Melkani}}, \bibinfo {author} {\bibfnamefont {C.}~\bibnamefont {Gneiting}}, \
  and\ \bibinfo {author} {\bibfnamefont {F.}~\bibnamefont {Nori}},\ }\href@noop
  {} {\bibfield  {journal} {\bibinfo  {journal} {Phys. Rev. A}\ }\textbf
  {\bibinfo {volume} {102}},\ \bibinfo {pages} {022412} (\bibinfo {year}
  {2020})}\BibitemShut {NoStop}%
\bibitem [{\citenamefont {Hsieh}\ \emph {et~al.}(2022)\citenamefont {Hsieh},
  \citenamefont {Ning}, \citenamefont {Chen}, \citenamefont {Wu}, \citenamefont
  {Chen}, \citenamefont {Wu},\ and\ \citenamefont {Lee}}]{hsieh2022direct}%
  \BibitemOpen
  \bibfield  {author} {\bibinfo {author} {\bibfnamefont {H.-Y.}\ \bibnamefont
  {Hsieh}}, \bibinfo {author} {\bibfnamefont {J.}~\bibnamefont {Ning}},
  \bibinfo {author} {\bibfnamefont {Y.-R.}\ \bibnamefont {Chen}}, \bibinfo
  {author} {\bibfnamefont {H.-C.}\ \bibnamefont {Wu}}, \bibinfo {author}
  {\bibfnamefont {H.~L.}\ \bibnamefont {Chen}}, \bibinfo {author}
  {\bibfnamefont {C.-M.}\ \bibnamefont {Wu}}, \ and\ \bibinfo {author}
  {\bibfnamefont {R.-K.}\ \bibnamefont {Lee}},\ }\href@noop {} {\bibfield
  {journal} {\bibinfo  {journal} {Symmetry}\ }\textbf {\bibinfo {volume}
  {14}},\ \bibinfo {pages} {874} (\bibinfo {year} {2022})}\BibitemShut
  {NoStop}%
\bibitem [{\citenamefont {Genois}\ \emph {et~al.}(2021)\citenamefont {Genois},
  \citenamefont {Gross}, \citenamefont {Di~Paolo}, \citenamefont {Stevenson},
  \citenamefont {Koolstra}, \citenamefont {Hashim}, \citenamefont {Siddiqi},\
  and\ \citenamefont {Blais}}]{genois2021quantum}%
  \BibitemOpen
  \bibfield  {author} {\bibinfo {author} {\bibfnamefont {{\'E}.}~\bibnamefont
  {Genois}}, \bibinfo {author} {\bibfnamefont {J.~A.}\ \bibnamefont {Gross}},
  \bibinfo {author} {\bibfnamefont {A.}~\bibnamefont {Di~Paolo}}, \bibinfo
  {author} {\bibfnamefont {N.~J.}\ \bibnamefont {Stevenson}}, \bibinfo {author}
  {\bibfnamefont {G.}~\bibnamefont {Koolstra}}, \bibinfo {author}
  {\bibfnamefont {A.}~\bibnamefont {Hashim}}, \bibinfo {author} {\bibfnamefont
  {I.}~\bibnamefont {Siddiqi}}, \ and\ \bibinfo {author} {\bibfnamefont
  {A.}~\bibnamefont {Blais}},\ }\href@noop {} {\bibfield  {journal} {\bibinfo
  {journal} {PRX Quantum}\ }\textbf {\bibinfo {volume} {2}},\ \bibinfo {pages}
  {040355} (\bibinfo {year} {2021})}\BibitemShut {NoStop}%
\bibitem [{\citenamefont {Teo}\ \emph {et~al.}(2021)\citenamefont {Teo},
  \citenamefont {Shin}, \citenamefont {Jeong}, \citenamefont {Kim},
  \citenamefont {Kim}, \citenamefont {Struchalin}, \citenamefont {Kovlakov},
  \citenamefont {Straupe}, \citenamefont {Kulik}, \citenamefont {Leuchs},\ and\
  \citenamefont {S{\'{a}}nchez-Soto}}]{teo2021benchmarking}%
  \BibitemOpen
  \bibfield  {author} {\bibinfo {author} {\bibfnamefont {Y.~S.}\ \bibnamefont
  {Teo}}, \bibinfo {author} {\bibfnamefont {S.}~\bibnamefont {Shin}}, \bibinfo
  {author} {\bibfnamefont {H.}~\bibnamefont {Jeong}}, \bibinfo {author}
  {\bibfnamefont {Y.}~\bibnamefont {Kim}}, \bibinfo {author} {\bibfnamefont
  {Y.-H.}\ \bibnamefont {Kim}}, \bibinfo {author} {\bibfnamefont {G.~I.}\
  \bibnamefont {Struchalin}}, \bibinfo {author} {\bibfnamefont {E.~V.}\
  \bibnamefont {Kovlakov}}, \bibinfo {author} {\bibfnamefont {S.~S.}\
  \bibnamefont {Straupe}}, \bibinfo {author} {\bibfnamefont {S.~P.}\
  \bibnamefont {Kulik}}, \bibinfo {author} {\bibfnamefont {G.}~\bibnamefont
  {Leuchs}}, \ and\ \bibinfo {author} {\bibfnamefont {L.~L.}\ \bibnamefont
  {S{\'{a}}nchez-Soto}},\ }\href {\doibase 10.1088/1367-2630/ac1fcb} {\bibfield
   {journal} {\bibinfo  {journal} {New J. Phys.}\ }\textbf {\bibinfo {volume}
  {23}},\ \bibinfo {pages} {103021} (\bibinfo {year} {2021})}\BibitemShut
  {NoStop}%
\bibitem [{\citenamefont {Tiunov}\ \emph {et~al.}(2020)\citenamefont {Tiunov},
  \citenamefont {Tiunova}, \citenamefont {Ulanov}, \citenamefont {Lvovsky},\
  and\ \citenamefont {Fedorov}}]{tiunov2020experimental}%
  \BibitemOpen
  \bibfield  {author} {\bibinfo {author} {\bibfnamefont {E.~S.}\ \bibnamefont
  {Tiunov}}, \bibinfo {author} {\bibfnamefont {V.}~\bibnamefont {Tiunova}},
  \bibinfo {author} {\bibfnamefont {A.~E.}\ \bibnamefont {Ulanov}}, \bibinfo
  {author} {\bibfnamefont {A.}~\bibnamefont {Lvovsky}}, \ and\ \bibinfo
  {author} {\bibfnamefont {A.}~\bibnamefont {Fedorov}},\ }\href@noop {}
  {\bibfield  {journal} {\bibinfo  {journal} {Optica}\ }\textbf {\bibinfo
  {volume} {7}},\ \bibinfo {pages} {448} (\bibinfo {year} {2020})}\BibitemShut
  {NoStop}%
\bibitem [{\citenamefont {Palmieri}\ \emph {et~al.}(2020)\citenamefont
  {Palmieri}, \citenamefont {Kovlakov}, \citenamefont {Bianchi}, \citenamefont
  {Yudin}, \citenamefont {Straupe}, \citenamefont {Biamonte},\ and\
  \citenamefont {Kulik}}]{palmieri2020experimental}%
  \BibitemOpen
  \bibfield  {author} {\bibinfo {author} {\bibfnamefont {A.~M.}\ \bibnamefont
  {Palmieri}}, \bibinfo {author} {\bibfnamefont {E.}~\bibnamefont {Kovlakov}},
  \bibinfo {author} {\bibfnamefont {F.}~\bibnamefont {Bianchi}}, \bibinfo
  {author} {\bibfnamefont {D.}~\bibnamefont {Yudin}}, \bibinfo {author}
  {\bibfnamefont {S.}~\bibnamefont {Straupe}}, \bibinfo {author} {\bibfnamefont
  {J.~D.}\ \bibnamefont {Biamonte}}, \ and\ \bibinfo {author} {\bibfnamefont
  {S.}~\bibnamefont {Kulik}},\ }\href@noop {} {\bibfield  {journal} {\bibinfo
  {journal} {npj Quantum Inf.}\ }\textbf {\bibinfo {volume} {6}},\ \bibinfo
  {pages} {20} (\bibinfo {year} {2020})}\BibitemShut {NoStop}%
\bibitem [{\citenamefont {Neugebauer}\ \emph {et~al.}(2020)\citenamefont
  {Neugebauer}, \citenamefont {Fischer}, \citenamefont {J{\"a}ger},
  \citenamefont {Czischek}, \citenamefont {Jochim}, \citenamefont
  {Weidem{\"u}ller},\ and\ \citenamefont
  {G{\"a}rttner}}]{neugebauer2020neural}%
  \BibitemOpen
  \bibfield  {author} {\bibinfo {author} {\bibfnamefont {M.}~\bibnamefont
  {Neugebauer}}, \bibinfo {author} {\bibfnamefont {L.}~\bibnamefont {Fischer}},
  \bibinfo {author} {\bibfnamefont {A.}~\bibnamefont {J{\"a}ger}}, \bibinfo
  {author} {\bibfnamefont {S.}~\bibnamefont {Czischek}}, \bibinfo {author}
  {\bibfnamefont {S.}~\bibnamefont {Jochim}}, \bibinfo {author} {\bibfnamefont
  {M.}~\bibnamefont {Weidem{\"u}ller}}, \ and\ \bibinfo {author} {\bibfnamefont
  {M.}~\bibnamefont {G{\"a}rttner}},\ }\href@noop {} {\bibfield  {journal}
  {\bibinfo  {journal} {Phys. Rev. A}\ }\textbf {\bibinfo {volume} {102}},\
  \bibinfo {pages} {042604} (\bibinfo {year} {2020})}\BibitemShut {NoStop}%
\bibitem [{\citenamefont {Wang}\ \emph {et~al.}(2021)\citenamefont {Wang},
  \citenamefont {Hernani-Morales}, \citenamefont {Mart{\'\i}n-Guerrero},
  \citenamefont {Solano},\ and\ \citenamefont
  {Albarr{\'a}n-Arriagada}}]{wang2021quantum}%
  \BibitemOpen
  \bibfield  {author} {\bibinfo {author} {\bibfnamefont {R.}~\bibnamefont
  {Wang}}, \bibinfo {author} {\bibfnamefont {C.}~\bibnamefont
  {Hernani-Morales}}, \bibinfo {author} {\bibfnamefont {J.~D.}\ \bibnamefont
  {Mart{\'\i}n-Guerrero}}, \bibinfo {author} {\bibfnamefont {E.}~\bibnamefont
  {Solano}}, \ and\ \bibinfo {author} {\bibfnamefont {F.}~\bibnamefont
  {Albarr{\'a}n-Arriagada}},\ }\href@noop {} {\bibfield  {journal} {\bibinfo
  {journal} {Quantum Sci. Technol.}\ }\textbf {\bibinfo {volume} {7}},\
  \bibinfo {pages} {015010} (\bibinfo {year} {2021})}\BibitemShut {NoStop}%
\bibitem [{\citenamefont {Ahmed}\ \emph
  {et~al.}(2021{\natexlab{b}})\citenamefont {Ahmed}, \citenamefont {Mu{\~n}oz},
  \citenamefont {Nori},\ and\ \citenamefont {Kockum}}]{ahmed2021quantum}%
  \BibitemOpen
  \bibfield  {author} {\bibinfo {author} {\bibfnamefont {S.}~\bibnamefont
  {Ahmed}}, \bibinfo {author} {\bibfnamefont {C.~S.}\ \bibnamefont
  {Mu{\~n}oz}}, \bibinfo {author} {\bibfnamefont {F.}~\bibnamefont {Nori}}, \
  and\ \bibinfo {author} {\bibfnamefont {A.~F.}\ \bibnamefont {Kockum}},\
  }\href@noop {} {\bibfield  {journal} {\bibinfo  {journal} {Phys. Rev. Lett.}\
  }\textbf {\bibinfo {volume} {127}},\ \bibinfo {pages} {140502} (\bibinfo
  {year} {2021}{\natexlab{b}})}\BibitemShut {NoStop}%
\bibitem [{\citenamefont {Carrasquilla}\ \emph {et~al.}(2019)\citenamefont
  {Carrasquilla}, \citenamefont {Torlai}, \citenamefont {Melko},\ and\
  \citenamefont {Aolita}}]{carrasquilla2019reconstructing}%
  \BibitemOpen
  \bibfield  {author} {\bibinfo {author} {\bibfnamefont {J.}~\bibnamefont
  {Carrasquilla}}, \bibinfo {author} {\bibfnamefont {G.}~\bibnamefont
  {Torlai}}, \bibinfo {author} {\bibfnamefont {R.~G.}\ \bibnamefont {Melko}}, \
  and\ \bibinfo {author} {\bibfnamefont {L.}~\bibnamefont {Aolita}},\
  }\href@noop {} {\bibfield  {journal} {\bibinfo  {journal} {Nat. Mach.
  Intell.}\ }\textbf {\bibinfo {volume} {1}},\ \bibinfo {pages} {155} (\bibinfo
  {year} {2019})}\BibitemShut {NoStop}%
\bibitem [{\citenamefont {Lohani}\ \emph
  {et~al.}(2020{\natexlab{b}})\citenamefont {Lohani}, \citenamefont {Knutson},\
  and\ \citenamefont {Glasser}}]{lohani2020generative}%
  \BibitemOpen
  \bibfield  {author} {\bibinfo {author} {\bibfnamefont {S.}~\bibnamefont
  {Lohani}}, \bibinfo {author} {\bibfnamefont {E.~M.}\ \bibnamefont {Knutson}},
  \ and\ \bibinfo {author} {\bibfnamefont {R.~T.}\ \bibnamefont {Glasser}},\
  }\href@noop {} {\bibfield  {journal} {\bibinfo  {journal} {Communications
  Physics}\ }\textbf {\bibinfo {volume} {3}},\ \bibinfo {pages} {1} (\bibinfo
  {year} {2020}{\natexlab{b}})}\BibitemShut {NoStop}%
\bibitem [{\citenamefont {Borah}\ \emph {et~al.}(2021)\citenamefont {Borah},
  \citenamefont {Sarma}, \citenamefont {Kewming}, \citenamefont {Milburn},\
  and\ \citenamefont {Twamley}}]{borah2021measurement}%
  \BibitemOpen
  \bibfield  {author} {\bibinfo {author} {\bibfnamefont {S.}~\bibnamefont
  {Borah}}, \bibinfo {author} {\bibfnamefont {B.}~\bibnamefont {Sarma}},
  \bibinfo {author} {\bibfnamefont {M.}~\bibnamefont {Kewming}}, \bibinfo
  {author} {\bibfnamefont {G.~J.}\ \bibnamefont {Milburn}}, \ and\ \bibinfo
  {author} {\bibfnamefont {J.}~\bibnamefont {Twamley}},\ }\href@noop {}
  {\bibfield  {journal} {\bibinfo  {journal} {Phys. Rev. Lett.}\ }\textbf
  {\bibinfo {volume} {127}},\ \bibinfo {pages} {190403} (\bibinfo {year}
  {2021})}\BibitemShut {NoStop}%
\bibitem [{\citenamefont {Gross}\ \emph {et~al.}(2010)\citenamefont {Gross},
  \citenamefont {Liu}, \citenamefont {Flammia}, \citenamefont {Becker},\ and\
  \citenamefont {Eisert}}]{gross2010quantum}%
  \BibitemOpen
  \bibfield  {author} {\bibinfo {author} {\bibfnamefont {D.}~\bibnamefont
  {Gross}}, \bibinfo {author} {\bibfnamefont {Y.-K.}\ \bibnamefont {Liu}},
  \bibinfo {author} {\bibfnamefont {S.~T.}\ \bibnamefont {Flammia}}, \bibinfo
  {author} {\bibfnamefont {S.}~\bibnamefont {Becker}}, \ and\ \bibinfo {author}
  {\bibfnamefont {J.}~\bibnamefont {Eisert}},\ }\href@noop {} {\bibfield
  {journal} {\bibinfo  {journal} {Phys. Rev. Lett.}\ }\textbf {\bibinfo
  {volume} {105}},\ \bibinfo {pages} {150401} (\bibinfo {year}
  {2010})}\BibitemShut {NoStop}%
\bibitem [{\citenamefont {H\"{a}ffner}\ \emph {et~al.}(2005)\citenamefont
  {H\"{a}ffner}, \citenamefont {H\"{a}nsel}, \citenamefont {Roos},
  \citenamefont {Benhelm}, \citenamefont {Chek-al kar}, \citenamefont
  {Chwalla}, \citenamefont {K\"{o}rber}, \citenamefont {Rapol}, \citenamefont
  {Riebe}, \citenamefont {Schmidt}, \citenamefont {Becher}, \citenamefont
  {G\"{u}hne}, \citenamefont {D\"{u}r},\ and\ \citenamefont
  {Blatt}}]{haffner2005scalable}%
  \BibitemOpen
  \bibfield  {author} {\bibinfo {author} {\bibfnamefont {H.}~\bibnamefont
  {H\"{a}ffner}}, \bibinfo {author} {\bibfnamefont {W.}~\bibnamefont
  {H\"{a}nsel}}, \bibinfo {author} {\bibfnamefont {C.~F.}\ \bibnamefont
  {Roos}}, \bibinfo {author} {\bibfnamefont {J.}~\bibnamefont {Benhelm}},
  \bibinfo {author} {\bibfnamefont {D.}~\bibnamefont {Chek-al kar}}, \bibinfo
  {author} {\bibfnamefont {M.}~\bibnamefont {Chwalla}}, \bibinfo {author}
  {\bibfnamefont {T.}~\bibnamefont {K\"{o}rber}}, \bibinfo {author}
  {\bibfnamefont {U.~D.}\ \bibnamefont {Rapol}}, \bibinfo {author}
  {\bibfnamefont {M.}~\bibnamefont {Riebe}}, \bibinfo {author} {\bibfnamefont
  {P.~O.}\ \bibnamefont {Schmidt}}, \bibinfo {author} {\bibfnamefont
  {C.}~\bibnamefont {Becher}}, \bibinfo {author} {\bibfnamefont
  {O.}~\bibnamefont {G\"{u}hne}}, \bibinfo {author} {\bibfnamefont
  {W.}~\bibnamefont {D\"{u}r}}, \ and\ \bibinfo {author} {\bibfnamefont
  {R.}~\bibnamefont {Blatt}},\ }\href {\doibase 10.1038/nature04279} {\bibfield
   {journal} {\bibinfo  {journal} {Nature}\ }\textbf {\bibinfo {volume}
  {438}},\ \bibinfo {pages} {643} (\bibinfo {year} {2005})}\BibitemShut
  {NoStop}%
\bibitem [{\citenamefont {Sommers}\ and\ \citenamefont
  {Zyczkowski}(2003)}]{sommers2003bures}%
  \BibitemOpen
  \bibfield  {author} {\bibinfo {author} {\bibfnamefont {H.-J.}\ \bibnamefont
  {Sommers}}\ and\ \bibinfo {author} {\bibfnamefont {K.}~\bibnamefont
  {Zyczkowski}},\ }\href@noop {} {\bibfield  {journal} {\bibinfo  {journal} {J.
  Phys. A: Math. Gen.}\ }\textbf {\bibinfo {volume} {36}},\ \bibinfo {pages}
  {10083} (\bibinfo {year} {2003})}\BibitemShut {NoStop}%
\bibitem [{\citenamefont {Zyczkowski}\ and\ \citenamefont
  {Sommers}(2003)}]{zyczkowski2003hilbert}%
  \BibitemOpen
  \bibfield  {author} {\bibinfo {author} {\bibfnamefont {K.}~\bibnamefont
  {Zyczkowski}}\ and\ \bibinfo {author} {\bibfnamefont {H.-J.}\ \bibnamefont
  {Sommers}},\ }\href@noop {} {\bibfield  {journal} {\bibinfo  {journal} {J.
  Phys. A: Math. Gen.}\ }\textbf {\bibinfo {volume} {36}},\ \bibinfo {pages}
  {10115} (\bibinfo {year} {2003})}\BibitemShut {NoStop}%
\bibitem [{\citenamefont {Al~Osipov}\ \emph {et~al.}(2010)\citenamefont
  {Al~Osipov}, \citenamefont {Sommers},\ and\ \citenamefont
  {{\.Z}yczkowski}}]{al2010random}%
  \BibitemOpen
  \bibfield  {author} {\bibinfo {author} {\bibfnamefont {V.}~\bibnamefont
  {Al~Osipov}}, \bibinfo {author} {\bibfnamefont {H.-J.}\ \bibnamefont
  {Sommers}}, \ and\ \bibinfo {author} {\bibfnamefont {K.}~\bibnamefont
  {{\.Z}yczkowski}},\ }\href@noop {} {\bibfield  {journal} {\bibinfo  {journal}
  {J. Phys. A: Math. Theor.}\ }\textbf {\bibinfo {volume} {43}},\ \bibinfo
  {pages} {055302} (\bibinfo {year} {2010})}\BibitemShut {NoStop}%
\bibitem [{LOH(2022)}]{LOHANI_machine-learning-for-physical-sciences_2022}%
  \BibitemOpen
  \href {https://github.com/slohani-ai/machine-learning-for-physical-sciences}
  {\enquote {\bibinfo {title}
  {https://github.com/slohani-ai/machine-learning-for-physical-sciences},}\ }
  (\bibinfo {year} {2022})\BibitemShut {NoStop}%
\bibitem [{\citenamefont {Nielsen}(1996)}]{nielsen1996entanglement}%
  \BibitemOpen
  \bibfield  {author} {\bibinfo {author} {\bibfnamefont {M.~A.}\ \bibnamefont
  {Nielsen}},\ }\href@noop {} {\bibfield  {journal} {\bibinfo  {journal}
  {arXiv:quant-ph/9606012}\ } (\bibinfo {year} {1996})}\BibitemShut {NoStop}%
\bibitem [{\citenamefont {Wilde}(2011)}]{wilde2011classical}%
  \BibitemOpen
  \bibfield  {author} {\bibinfo {author} {\bibfnamefont {M.~M.}\ \bibnamefont
  {Wilde}},\ }\href@noop {} {\bibfield  {journal} {\bibinfo  {journal}
  {arXiv:1106.1445}\ } (\bibinfo {year} {2011})}\BibitemShut {NoStop}%
\bibitem [{\citenamefont {{\.Z}yczkowski}\ and\ \citenamefont
  {Sommers}(2005)}]{zyczkowski2005average}%
  \BibitemOpen
  \bibfield  {author} {\bibinfo {author} {\bibfnamefont {K.}~\bibnamefont
  {{\.Z}yczkowski}}\ and\ \bibinfo {author} {\bibfnamefont {H.-J.}\
  \bibnamefont {Sommers}},\ }\href@noop {} {\bibfield  {journal} {\bibinfo
  {journal} {Phys. Rev. A}\ }\textbf {\bibinfo {volume} {71}},\ \bibinfo
  {pages} {032313} (\bibinfo {year} {2005})}\BibitemShut {NoStop}%
\end{thebibliography}%

\end{document}